\documentclass[12pt,preprint]{aastex}

\begin{document}
 
 \title{A High Precision, Optical Polarimeter to Measure Inclinations of High Mass X-Ray Binaries}
 
 \author{Sloane J. Wiktorowicz\altaffilmark{1} \& Keith Matthews\altaffilmark{2}}
 \email{sloane@berkeley.edu}
 
 \altaffiltext{1}{Planetary Science Department, California Institute of Technology, Pasadena, CA 91125 \\
 Current address: Astronomy Department, University of California, Berkeley, CA 94720}
 \altaffiltext{2}{Physics Department, California Institute of Technology, Pasadena, CA 91125}
  
  \begin{abstract}
We present commissioning data for the POLISH instrument obtained on the Hale 5-m telescope. The goal of this high precision polarimeter is to constrain orbital inclination of high mass X-ray binaries and to therefore obtain independent mass estimates for their black hole companions. We have obtained photon shot noise limited precision on standard stars, and we have measured the polarization of bright stars at the part per million level on a nightly basis. Systematic effects have been reduced to less than $1\%$ of the measured polarization for polarized sources and to the part per million level for weakly polarized sources. The high sensitivity of this instrument to asymmetry suggests that valuable contributions will be made in many other fields, including studies of extrasolar planets, debris disks, and stellar astrophysics.
\end{abstract}


\section{Introduction}
 
While most astrophysical objects require many parameters in order to be fully described, black holes are unique in that only three parameters suffice: mass, spin, and charge. Mass and spin describe the black hole's gravitational field and event horizon location. Therefore, black holes provide a rare opportunity for theory and observation to jointly pursue a few quantities to completely describe one of the most exotic kinds of objects in the Universe.\\

Though modeling of the X-ray spectral energy distribution is somewhat effective in constraining black hole spin (McClintock et al. 2006), important constraints on black hole mass exist in the case of high mass X-ray binaries (hereafter HMXBs). These binaries consist of an O or B type supergiant and a black hole or neutron star. The most well-studied of these, Cygnus X-1, is thought to consist of a $40 \pm 10$ $M_\sun$, O9.7Iab star and a $13.5-29$ $M_\sun$ black hole at a distance of $2.2 \pm 0.2$ kpc (Zi\'{o}lkowski 2005). While the constraints on the mass of the compact object are tight enough to declare that it is a black hole, they are insufficient to permit precise modeling of the progenitor star's mass. We have commissioned a polarimeter on the Hale 5-m telescope at Palomar Observatory in California to provide an independent method for determining black hole mass. This polarimeter has the potential to constrain the mass of the Cygnus X-1 black hole to a few solar masses. \\

\section{Black Hole Mass from Polarimetry}
 
 A wealth of radial velocity data exists for Cygnus X-1 (Gies et al. 2003) and other HMXBs. However, in the same way that precise masses are elusive for non-transiting extrasolar planets, determination of precise black hole mass is hindered by unknown orbital inclination. This is evidenced by the mass function, which is the end product of radial velocity observations. For Cygnus X-1, Gies et al. (2003) quote the following value for the mass function:

\begin{equation}
f\left(M_{\rm X}\right)=\frac{M_{\rm opt}\sin^3i}{q\left(1+q\right)^2}=0.251\pm0.007 \, M_{\sun}
\end{equation}
\bigskip

\noindent Here, $M_X$ is the mass of the black hole, $M_{\rm{opt}}$ is the mass of the visible binary component, $i$ is the system inclination, and $q$ is the mass ratio of the visible component to the black hole. Thus, an observational technique that constrains orbital inclination can take advantage of radial velocity data and offer an estimate of black hole mass.\\

\begin{figure}[t]     
 \begin{center}
  \scalebox{0.4}{\includegraphics{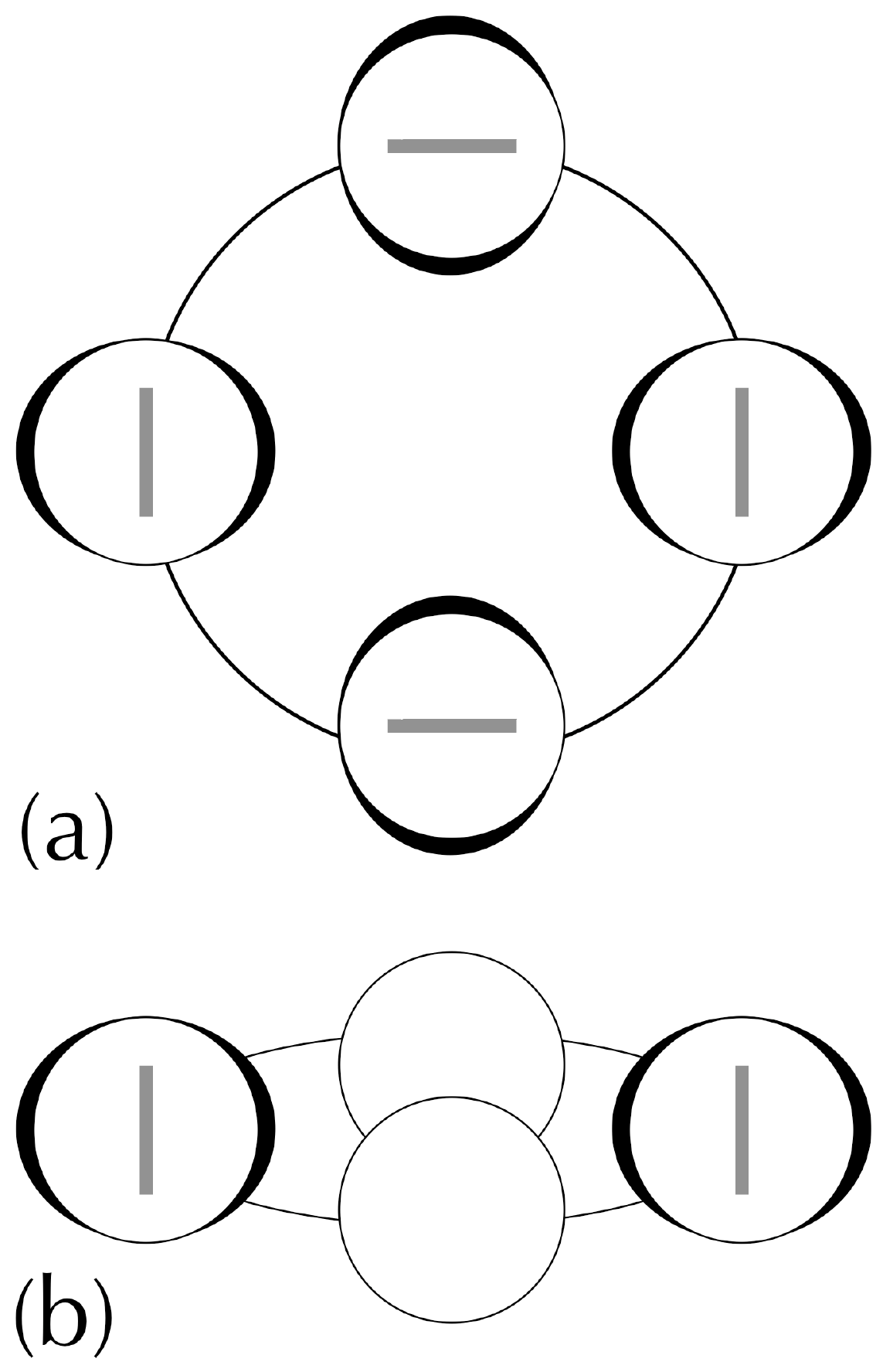}}
  \end{center}
 \caption[Orbital modulation of system polarization for HMXBs]{Orbital modulation of system polarization for HMXBs. The degree of polarization is represented by the black tidal bulges (the exact cause of polarization is irrelevant to this figure), and position angle of net polarization is given by the orientation of the grey lines. The face-on case is shown in (a), and the edge-on case is shown in (b). The circumbinary envelopes have been drawn displaced from the center of mass for clarity.}   
 \label{2myfiga}        
\end{figure}

Since system polarization is a geometric effect, the polarization of an HMXB system can be used to determine geometric information about the system, such as orbital inclination. The effective temperature of the supergiant in an HMXB is $T_{\rm eff} \approx$ 30,000 K, which is hot enough to ionize photospheric hydrogen. This generates a high density of free electrons that Thomson-scatter emitted light from the supergiant. While net linear polarization from a spherical cloud of free electrons is zero, asymmetry in the system causes net polarization (Brown \& McLean 1977). The tidal effects of the black hole cause such an asymmetry in the circumbinary envelope, and the orbital modulation of polarization is the key to determining orbital inclination. For instance, consider a face-on HMXB with zero eccentricity and an optically thin circumbinary envelope (Figure \ref{2myfiga}a). The total amount of observed, polarized light is independent of orbital phase, and the degree of polarization is therefore constant. However, the angle of net polarization, which is perpendicular to the scattering plane, rotates as the binary progresses in its orbit.\\

In contrast, for a nearly edge-on geometry the degree of polarization varies significantly throughout the orbit, while the angle of net polarization is roughly constant (Figure \ref{2myfiga}b). Therefore, the modulation of the degree and angle of net polarization is a unique measure of orbital inclination for synchronously rotating HMXBs. Combining Equations 1 and 2 from Friend \& Cassinelli (1986) and Equation 2 from Brown et al. (2000), the polarization of an optically thin, axisymmetric envelope due to Thomson scattering is

\begin{equation}
P=\frac{3}{16}\sigma_T\Big(1-\cos^2\phi_{\rm orb}\sin^2i \Big)\int_{r_1}^{r_2}\! \! \! \int_{\mu_1}^{\mu_2}n_e\left(r\right)\left(1-\frac{R^2}{r^2}\right)^{1/2} \Big(1-3\mu^2 \Big)drd\mu
\end{equation}
\smallskip

\noindent Here, $\sigma_T$ is the Thomson scattering cross section, $\mu$ is the cosine of the polar angle, the electron number density as a function of radial coordinate $r$ is $n_e\left(r\right)=n_0 R^2/r^2$, $n_0$ is a scaling constant, stellar radius is $R$, system inclination is $i$, and orbital phase in radians is $\phi_{\rm orb}$. Two polarization periods occur per orbital period because of the $\cos^2\phi_{\rm orb}$ term.\\

This technique has been utilized by a few groups (Kemp et al. 1979; Dolan \& Tapia 1989; Wolinski et al. 1996), and Cygnus X-1 has been found to have variable polarization of order $\Delta P \approx 0.1\%$ with respect to its unpolarized flux. However, measurement precision from the above groups is of order one part in $10^4$. The derived inclination estimates were questioned by the community on the grounds that significant underestimation of error occurred because of limited measurement precision (Milgrom 1979; Aspin et al. 1981). Measuring inclination to $5^\circ$ requires polarimetric precision of one part in $10^{4}$ to one part in $10^{7}$, depending on system inclination \citep{asp81}. This requires at least $10^8$ to $10^{14}$ detected photons, which necessitates the use of large telescopes. Since our system combines a high precision instrument with a 5-m telescope, we aim to measure the polarimetric variability of Cygnus X-1 to better than one part in $10^4$.\\

\section{The POLISH Instrument}
 
\begin{figure}[p]     
 \begin{center}
  \scalebox{0.7}{\includegraphics{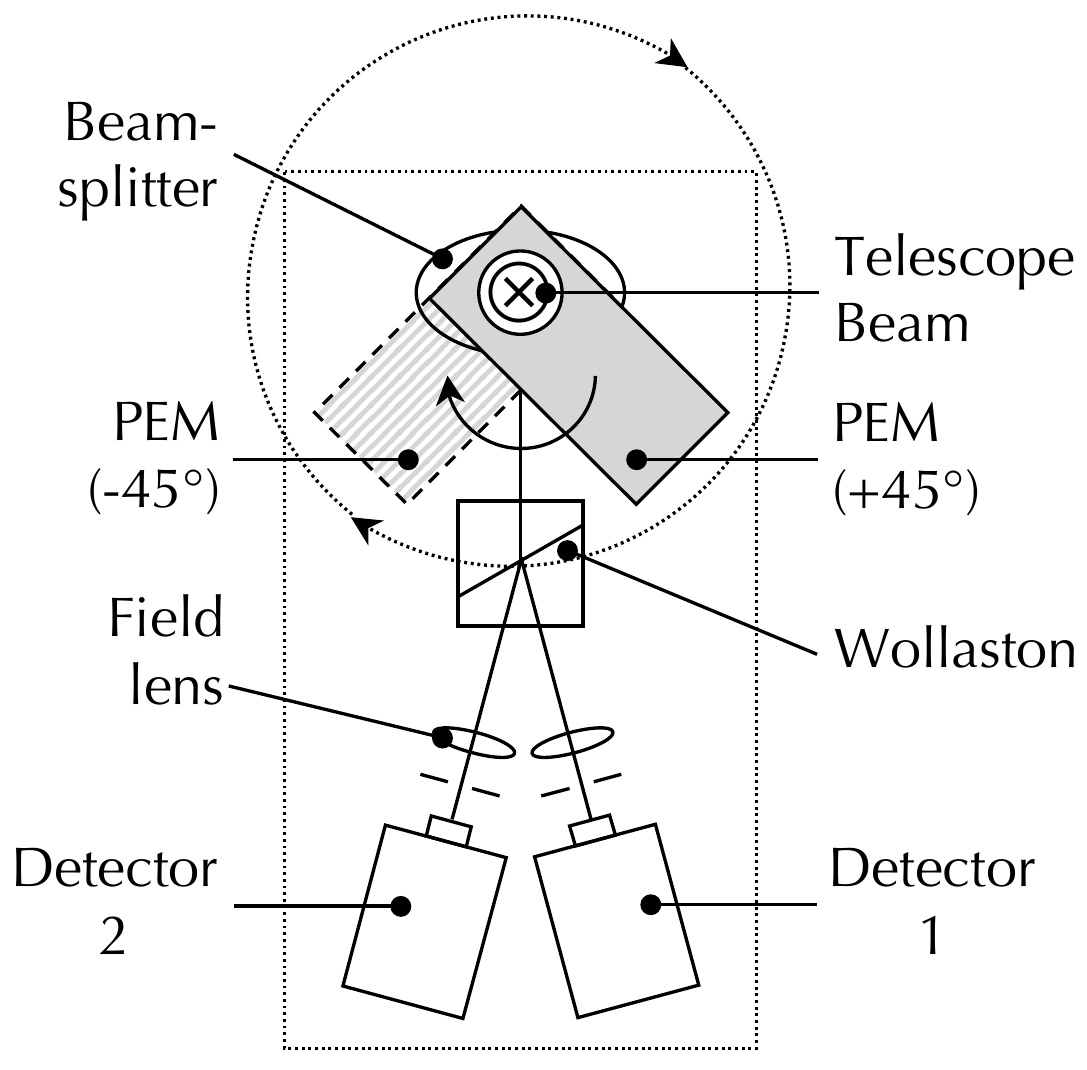}}
  \end{center}
 \caption[Plan view of the POLISH optical path]{Plan view of the POLISH optical path. The telescope beam is directed into the page through the center of the PEM aperture (the ``X" in the figure). The PEM is rotated to $\theta_{\rm PEM} = \pm 45^\circ$ with respect to the centerline, and the instrument itself (dotted box) can be independently rotated on the telescope through $\Delta\phi = 360^\circ$. Field stops are located between the field lenses and detectors.}   
 \label{2myfigb}        
\end{figure}

\begin{figure}[p]     
 \begin{center}
  \scalebox{0.55}{\includegraphics{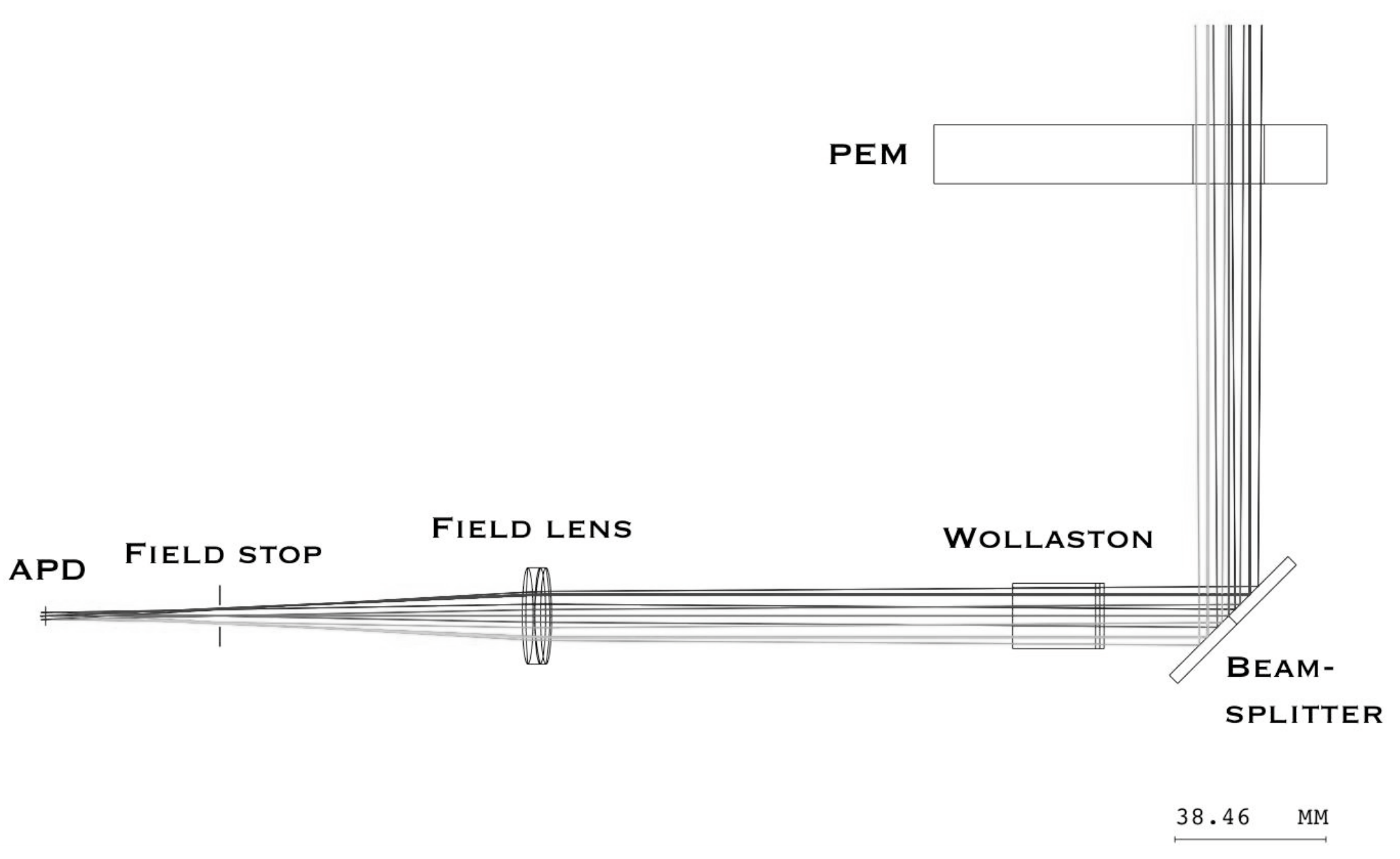}}
  \end{center}
 \caption[Ray trace diagram of the POLISH optical path]{Ray trace diagram of the POLISH optical path. The telescope beam enters the instrument from the top right of the figure.}   
 \label{2myfigc}        
\end{figure}

\begin{figure}[t]     
 \begin{center}
  \scalebox{0.8}{\includegraphics{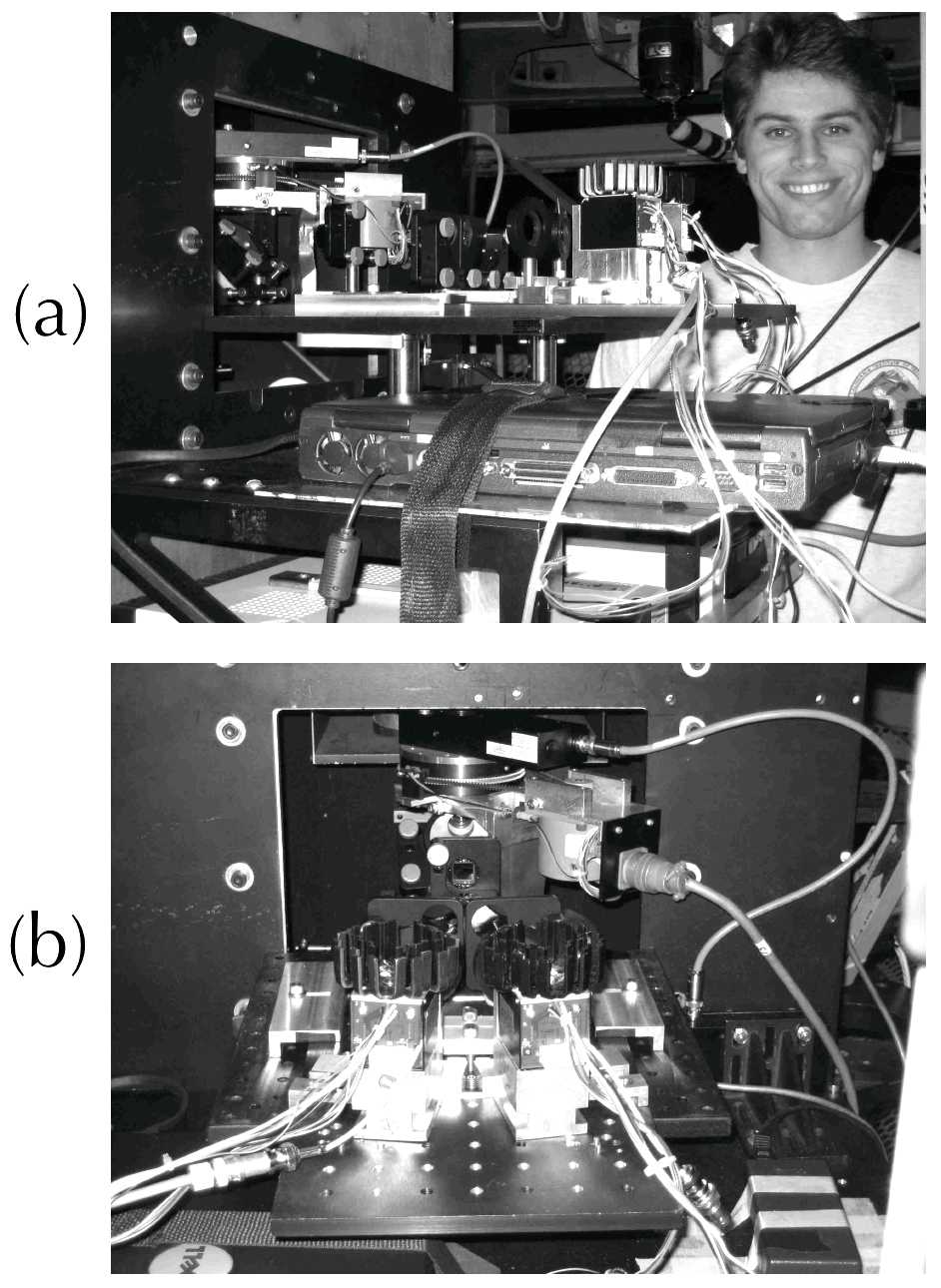}}
  \end{center}
 \caption[Photographs of the POLISH instrument]{Photographs of the POLISH instrument mounted at the Cassegrain focus of the Hale 5-m telescope.}   
 \label{2myfigz}        
\end{figure}

Polarimeters require the following fundamental components: a polarization modulator, analyzer, detector, and demodulator. The modulator induces a known, periodic characteristic to the unknown polarization of the input beam. The analyzer converts modulation in polarization to modulation in the beam's intensity, since most detectors are sensitive to intensity and not polarization. Finally, the demodulator extracts the component of the detector's output that varies at the known frequency of the modulator to reject noise. See Figure \ref{2myfigb} for a block diagram, Figure \ref{2myfigc} for a ray trace diagram, and Figure \ref{2myfigz} for photographs of the instrument. \\

Traditionally, the modulator is a rotating halfwave plate that rotates the linear polarization of the incident beam. The highest modulation frequency attainable with this component is of order 100 Hz, which is not fast enough to freeze out atmospheric turbulence or to mitigate electronic $1/f$ noise. Additionally, spatial inhomogeneity in both the retardance and cleanliness of the plate can introduce spurious signals, because the beam samples different sections of the halfwave plate at different times. The polarization goal for our instrument, POLISH (POLarimeter for Inclination Studies of High mass x-ray binaries/Hot jupiters), is one part per million on bright, unpolarized stars, which necessitates the use of a photoelastic modulator (hereafter PEM, Kemp 1969).\\

A PEM is a bar of optical materials (fused silica for use in optical light) in which a resonant acoustic signal at a frequency of tens of kilohertz is induced. The stress-birefringent property of the silica causes time-varying, sinusoidal retardance of the component of polarization oriented at $\pm 45^\circ$ with respect to the compression/extension axis, while the polarization components at $0^\circ$ and $90^\circ$ experience no retardance. that is, a PEM at $\theta_{\rm PEM} = \pm 45^\circ$ with respect to the Stokes $Q$ axis will cause retardance to the $Q/I$ Stokes parameter but not to the $U/I$ parameter. Since peak retardance is a function of the amplitude of the acoustic signal, both peak retardance and wavelength of peak retardance can be easily modified. We set the retardance amplitude to $\beta_0 = 0.383$ waves at 500 nm wavelength to give uniform PEM efficiency for both low and high linear polarizations. This also causes linear polarization to be directly proportional to the amplitude of the AC signal divided by the mean DC signal, which is derived in Appendix A.\\

The high frequency PEM modulation strongly reduces both atmospheric turbulence and electronic $1/f$ noise. Additionally, the beam always samples the same optical path during the modulation, because the strain on the modulator element is only of order ten parts per million (Kemp 1969). Operationally, a PEM is the opposite of a rotating halfwave plate: while the plate has a constant retardance and time-variable fast axis orientation, a PEM has constant fast axis orientation but time-variable retardance. Since the absolute value of the PEM's retardance determines the polarization of the beam at any instant, compression and extension of the fused silica bar affect the beam identically. Therefore, linear polarization is modulated at twice the frequency of the PEM modulation. We use the I/FS50 PEM and PEM100 controller from Hinds Instruments, Inc. The modulation frequency of this PEM is 50 kHz, and modulation of linearly polarized light occurs at 100 kHz.\\
 
Directly downstream from the PEM is a 95R/5T beamsplitter at $45^\circ$ incidence that allows $\approx 5\%$ of the stellar flux to fall on a Xybion CCD camera for on-axis guiding, while the remaining $\approx 95\%$ is reflected into a Wollaston prism toward the detectors. The beamsplitter has a 50-mm diameter, fused silica substrate from Edmund Optics with a custom 400 to 700 nm coating from Opticorp, Inc. The surface accuracy on the substrate is one-tenth of a wavelength.\\
 
We utilize a two-wedge, calcite Wollaston prism from Karl Lambrecht, Inc., as the analyzer. This prism separates each component of a single Stokes parameter into two beams. That is, the $+Q/I$ (or $+U/I$) component is split into one beam, and the $-Q/I$ (or $-U/I$) component is in the other beam. Both beams have equal deviation of $7.5^\circ$ from the optical axis, which allows the optical layout to be symmetric with respect to the optical axis. A two-wedge prism is used because the larger beam deviation of a three-wedge design would cause the instrument package to be larger than necessary. The surfaces of the Wollaston prism have an antireflection coating in the wavelength range of 400 to 700 nm, and the transmission in $V$ band is $\approx 97\%$ per surface. By injecting light through a linear polarizer with known fast axis orientation and then through the Wollaston prism, we find that the left beam seen from the detectors is vertically polarized ($-Q/I$ when projected onto the sky for a Cassegrain ring angle of $0^\circ$). The right beam is horizontally polarized ($+Q/I$ at $0^\circ$ ring angle).\\
 
Each Wollaston beam then impinges on an $f/5.6$, MgF$_2$ antireflection-coated field lens from Melles Griot. These lenses image the telescope secondary mirror onto the detectors, and they ensure starlight is uniformly spread over the detector active area even in the presence of image wander. Field stops are located in the image plane, after the field lenses, but these are not currently used because contamination of stellar polarization from the sky field is not significant. The beams reach the detectors with a diameter of $\approx 3$ mm, which underfills all detectors.\\

Since Cygnus X-1 has $V$ magnitude $\approx 9$, but the polarization standard stars observed can be as bright as $V \approx 3$, POLISH has two interchangeable pairs of detectors. Stars fainter than $V \approx 7$ are detected at a higher signal to noise ratio with the pair of Hamamatsu H9307-04 photomultiplier tubes (PMTs), while objects brighter than this will destroy the PMTs. Brighter stars are observed with custom-made Advanced Photonix SD197-70-72-661 (red enhanced) and SD197-70-74-661 (blue enhanced) avalanche photodiode modules (APDs). The high quantum efficiency of APDs is desirable on bright stars to minimize photon shot noise, while the low dark current of PMTs is desirable on fainter stars to minimize detector noise.\\
 
The detectors are not downstream from spectral filters, and they are integrated light detectors in both spatial and spectral senses. Spatial resolution is unnecessary, as the angular size of the Cygnus X-1 system is much smaller than the atmospheric seeing disk. Spectral resolution, while desirable, would seriously degrade the precision attainable with this instrument. Such resolution must be left for future generations of POLISH. Neglecting detector quantum efficiency, the instrumental throughput is calculated to be $74\%$, $77\%$, $58\%$, and $23\%$ in $B$, $V$, $R$, and $I$ bands. The throughput of the telescope/instrument system, again neglecting detector response, is calculated to be $60\%$, $62\%$, $47\%$, and $19\%$ in those bands.\\
 
The PMTs are identical, side-on modules with active area dimensions $3.7 \times 13.0$ mm. Their quantum efficiencies are quoted from the manufacturer to be $18\%$, $15\%$, $7\%$, and $0\%$ in $B$, $V$, $R$, and $I$ bands. The FWHM of the PMT bandpass is 475 nm (200 to 675 nm), but atmospheric transparency is of course low shortward of about 400 nm. Thus, the PMTs detect broadband $BV$ light. The PMT gain can be set by a potentiometer, and we set this gain to $G = 5\times10^6$ for all observations. The modules also have a $B = 200$ kHz bandwidth amplifier with transimpedance $T_A = 10^5$ V/A. The quoted output noise voltage resulting from dark current is $\sigma'_V = 10$ (typical) to $100$ $\mu$V (maximum), which implies a noise equivalent power of $\rm{NEP} = 0.04$ to 0.13 fW$/\sqrt{\rm{Hz}}$. Dark current is $i'_d = 0.1$ nA.\\
 
\begin{figure}[t]     
 \begin{center}
  \scalebox{0.6}{\includegraphics{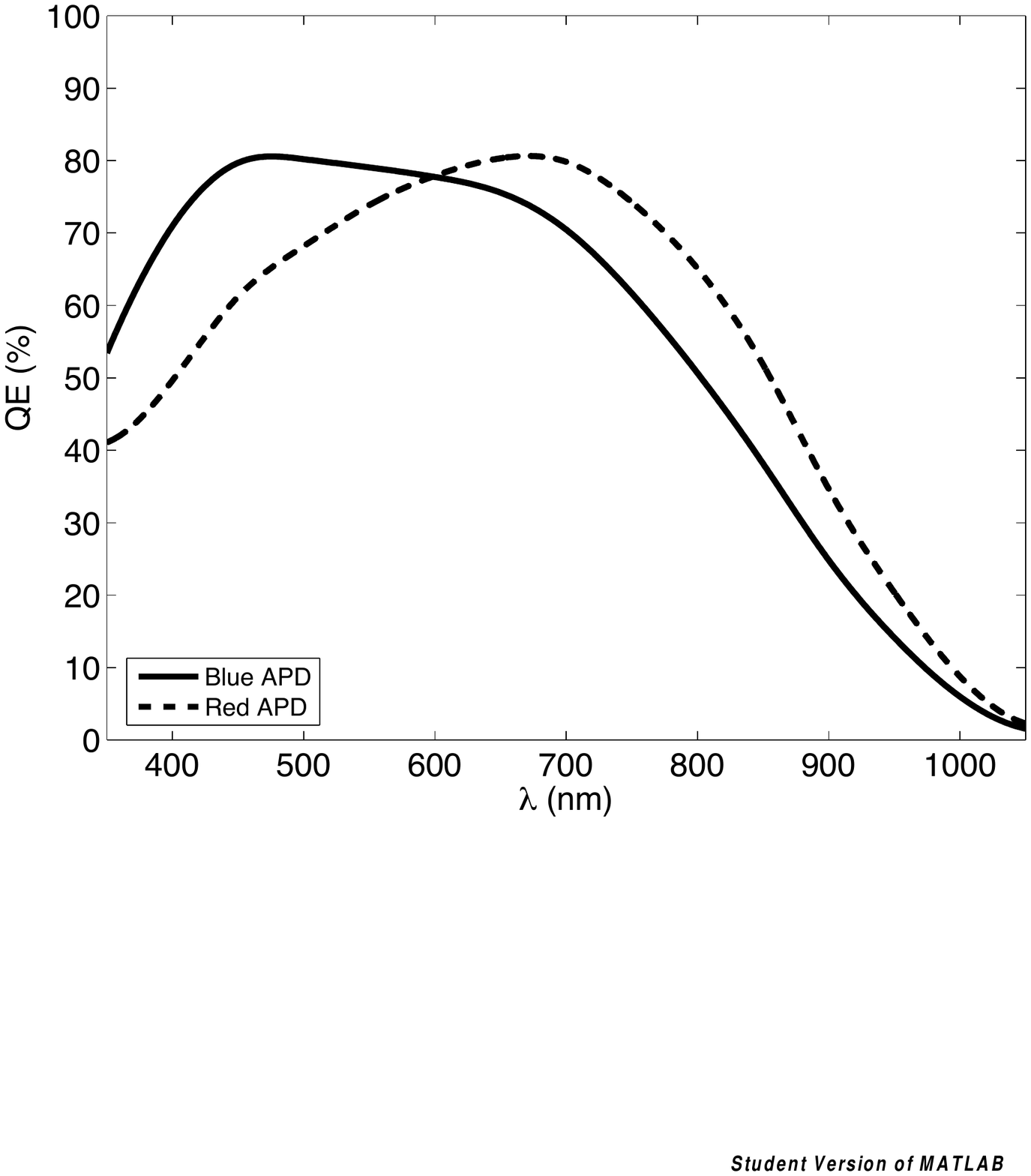}}
  \end{center}
 \caption[Quantum efficiency curves for the red enhanced and blue enhanced APDs]{Quantum efficiency curves for the red enhanced and blue enhanced APDs (detector 2 and 1, respectively).}   
 \label{2myfigq}        
\end{figure}

The APDs are optimized for 100 kHz operation and have 5 mm diameter, circular active areas. The APDs are not identical, as one beam is sampled by the red enhanced module while the other is sampled by the blue enhanced one. The quantum efficiencies for the red enhanced module are quoted as $58\%$, $73\%$, $78\%$, and $46\%$ in $B$, $V$, $R$, and $I$ bands (Figure \ref{2myfigq}). The FWHM of the bandpass for the red enhanced APD is 550 nm (350 to 900 nm), which corresponds to broadband $BVRI$ light. The blue enhanced module is quoted to have $75\%$, $82\%$, $67\%$, and $35\%$ quantum efficiencies. The FWHM of the bandpass for the blue enhanced APD is about 500 nm (350 to 850 nm), which corresponds to broadband $BVR$ light. The red module operates with an observed gain of $G = 220$, and the blue module operates at a quoted gain of $G = 300$. Transimpedance is $T_A = 4\times10^6$ V/A for both modules, and amplifier bandwidth is $B = 100$ kHz for the red enhanced and $B = 90$ kHz for the blue enhanced modules. After the APD chip is thermoelectrically cooled to $0^\circ$ C, dark current is measured to be $i'_d = 4.5$ nA and 3.5 nA at the output of the red and blue modules, respectively. Therefore, the noise equivalent power for each module is $\rm{NEP} = 39$ fW$/\sqrt{\rm{Hz}}$ and 9.7 fW$/\sqrt{\rm{Hz}}$, respectively. Each detector is supplied $\pm 12$V and $+5$V, and a 12V case fan blows heat from the APD heat sinks. \\
  
The demodulator picks out the component of the detected signal that varies at the reference frequency and ideally rejects signals at all other frequencies. The demodulator can either be software or hardware; POLISH makes use of one Stanford Research SR830 digital, dual-phase lock-in amplifier for each detector. The PEM controller sends a square wave reference signal to the lock-in amplifiers at twice the frequency of the PEM modulation, and each lock-in amplifier recovers $X$ (in phase with reference signal) and $Y$ ($90^\circ$ out of phase with the reference signal) components of the detector signal. Together, $X$ and $Y$ determine amplitude $R$ and phase $\Phi$ according to the following:

 \begin{mathletters}
\begin{equation}
R=\sqrt{2\left(X^2+Y^2\right)}
\label{2eqk}
\end{equation}
\begin{equation}
\Phi=\frac{1}{2} \arctan\left(Y,X\right)
\label{2eql}
\end{equation}
\end{mathletters}
 \smallskip

\noindent The lock-in amplifiers record the RMS components of the in-phase and quadrature phase signals, so multiplication by a factor of $\sqrt{2}$ is necessary to determine the amplitude of the AC signal. The notation of the argument of the arctangent is meant to account for the signs of $X$ and $Y$ when determining phase.\\

Signal phase allows direct measurement of the sign of each Stokes parameter (e.g., $+Q/I$ versus $-Q/I$). This is important, because insensitivity to sign would preclude direct measurement of more than $90^\circ$ of rotation of the Cygnus X-1 system. Both AC and DC signals from the detector must be recorded to measure polarization (Appendix A). The AC signals are recorded by the lock-in amplifiers, and each detector's DC signal is recorded by a separate HP 34401A digital voltmeter. To reject 60 Hz noise and its harmonics, each DC reading by the voltmeters consists of an integration over 10 power line cycles. Thus, the voltmeters record data at 6 Hz. The lock-in amplifiers may only sample the AC signal at powers of two in frequency, so we choose to record the AC data at 8 Hz. The discrepancy in sampling rates between AC and DC data is not important, because AC data should be normalized by mean DC data and not in a point-by-point fashion.\\

In order to Nyquist sample the AC data, we set the lock-in time constants to 30 ms. This gives an effective noise bandwidth of ENBW $= 2.6$ Hz. Therefore, we sample the AC data $8/2.6 \approx 3.1$ times per effective time constant, which satisfies the Nyquist criterion and reduces aliasing. The lock-in amplifiers thus measure the components of the AC signal with frequencies bounded by $f_0\left(1\pm1.3\times10^{-5}\right)$, where $f_0$ is the reference frequency in Hz. The auxiliary DC output of one lock-in amplifier is connected to a chopping motor on the telescope secondary mirror, which allows the secondary mirror to chop north to a sky field for sky subtraction of both AC and DC data.\\

POLISH is located at Cassegrain focus to ensure beam reflections of about $180^\circ$, and it samples the $f/72$ beam. These qualities minimize telescope polarization. To minimize instrument polarization, the PEM is the first instrumental optic encountered by the beam. The lock-in amplifiers and voltmeters are controlled by a laptop, which is mounted to the instrument, via the GPIB interface. Matlab R2006a from The MathWorks, Inc., is used to control the voltmeters and lock-in amplifiers, chop the secondary mirror, and rotate the Cassegrain ring to allow access to both linear Stokes parameters.\\

\section{Observing Strategy}
 
A similar, albeit larger, instrument named PlanetPol is mounted on the 4.2-m William Herschel Telescope in La Palma, Spain (Hough et al. 2006, hereafter HLB06). The goal of this instrument is to detect the modulation of linear polarization caused by stellar flux scattered by hot Jupiters. This observation requires polarimetric precision of one part per million to one part in ten million (Seager et al. 2000, Stam et al. 2004), which is a precision nearly achievable with PlanetPol on the brightest stars. We observed many of the polarized and unpolarized standard stars from HLB06 in addition to others from the combined polarimetric catalog of Heiles (2000). A list of the stars observed is given in Table \ref{2tablec}, and polarization values in parenthesis represent the $1\sigma$ uncertainty in the last two digits of the mean value. Polarization reference is given in the ``Ref" column. $V$ band magnitude and spectral type of the $\delta$ Cepheid star HD 187929 are from Bastien et al. (1988) and Oke (1961), respectively. The spectral type of HD 212311 is from Schmidt et al. (1992). All other positional and spectral information is from the SIMBAD database. Observations of Cygnus X-1 itself will be given in a forthcoming paper.\\

\begin{deluxetable}{c c c c c c c c c}
\tabletypesize{\scriptsize}
\tablecaption{Observed Standard Stars}
\tablewidth{0pt}
\tablehead{
\colhead{Name} & \colhead{Alt. Name} & \colhead{RA} & \colhead{Dec} & \colhead{$P$} & \colhead{$\Theta$} & \colhead{Ref} &  \colhead{$V$} & \colhead{Type}\\
& & & & (\%) & ($^\circ$) & & & }
\startdata
Algenib\tablenotemark{a} & $\gamma$ Peg & 00 13 14.23 & +15 11 00.9 & $0.0630(10)$ & 118.1(5) & 1 & 2.83 & B2IV\\
HD 7927 & $\phi$ Cas & 01 20 04.92 & +58 13 53.8 & $3.232(53)$ & 94.0(5) & 1 & 5.01 & F0Ia\\
HD 9270 & $\eta$ Psc & 01 31 29.07 & +15 20 44.8 & $0.0060(30)$ & 158(14) & 1 & 3.63 & G7IIa\\
HR 5854 & $\alpha$ Ser & 15 44 16.07 & +06 25 32.3 & $0.00043(10)$ & $-$ & 2 & 2.64 & K2IIIb\\
HD 147084 & $o$ Sco & 16 20 38.18 & $-$24 10 09.6 & $3.490(35)$ & 32.1(3) & 1 & 4.55 & A4II/III\\
HD 154445 & SAO 141513 & 17 05 32.24 & $-$00 53 31.7&  $3.420(24)$ & 90.2(2) & 1 & 5.64 & B1V\\
u Her\tablenotemark{b} & HD 156633 & 17 17 19.57 & +33 06 00.4 & $0.0(2)$ & $-$ & 1 & 4.80 & B1.5Vp+\\
$\gamma$ Oph\tablenotemark{c} & HD 161868 & 17 47 53.56 & +02 42 26.3 & $0.0080(10)$ & 33.3(3.6) & 1 & 3.75 & A0V\\
HD 157999 & $\sigma$ Oph & 17 26 30.98 & +04 08 25.1 & $1.010(35)$ & 85.9(1.0) & 1 & 4.34 & K3Iab\\
HD 187929\tablenotemark{d} & $\eta$ Aql & 19 52 28.37 & +01 00 20.4 & $1.685(3)$ & 94.2(1) & 2 & $3.5-4.3$ & (F6.5$-$G2)Ib\\
HD 204827 & SAO 33461 & 21 28 57.70 & +58 44 24.0 & $5.44(20)$ & 59.0(1.1) & 1 & 8.00 &O9.5V\\
HD 212311 & SAO 34361 & 22 21 58.55 & +56 31 52.8 & $0.02(5)$ & $-$ & 1 & 8.12 & A0V\\
HR 8974\tablenotemark{e} & $\gamma$ Cep & 23 39 20.85 & +77 37 56.2 & $0.00052(22)$ & $-$ & 2 & 3.23 & K1IV\\
\enddata
\label{2tablec}
\tablenotetext{a}{$\beta$ Cepheid, pulsator}
\tablenotetext{b}{$\beta$ Lyrid, eclipsing binary}
\tablenotetext{c}{Debris disk}
\tablenotetext{d}{$\delta$ Cepheid, pulsator}
\tablenotetext{e}{Extrasolar planet host}
\tablerefs{(1) Heiles (2000), (2) HLB06}
\tablecomments{Heiles (2000) polarization data are quoted for $V$ band, and HLB06 polarizations are measured in the wavelength range of 625 to 950 nm.}
\end{deluxetable}

\begin{figure}[p]     
 \begin{center}
  \scalebox{0.55}{\includegraphics{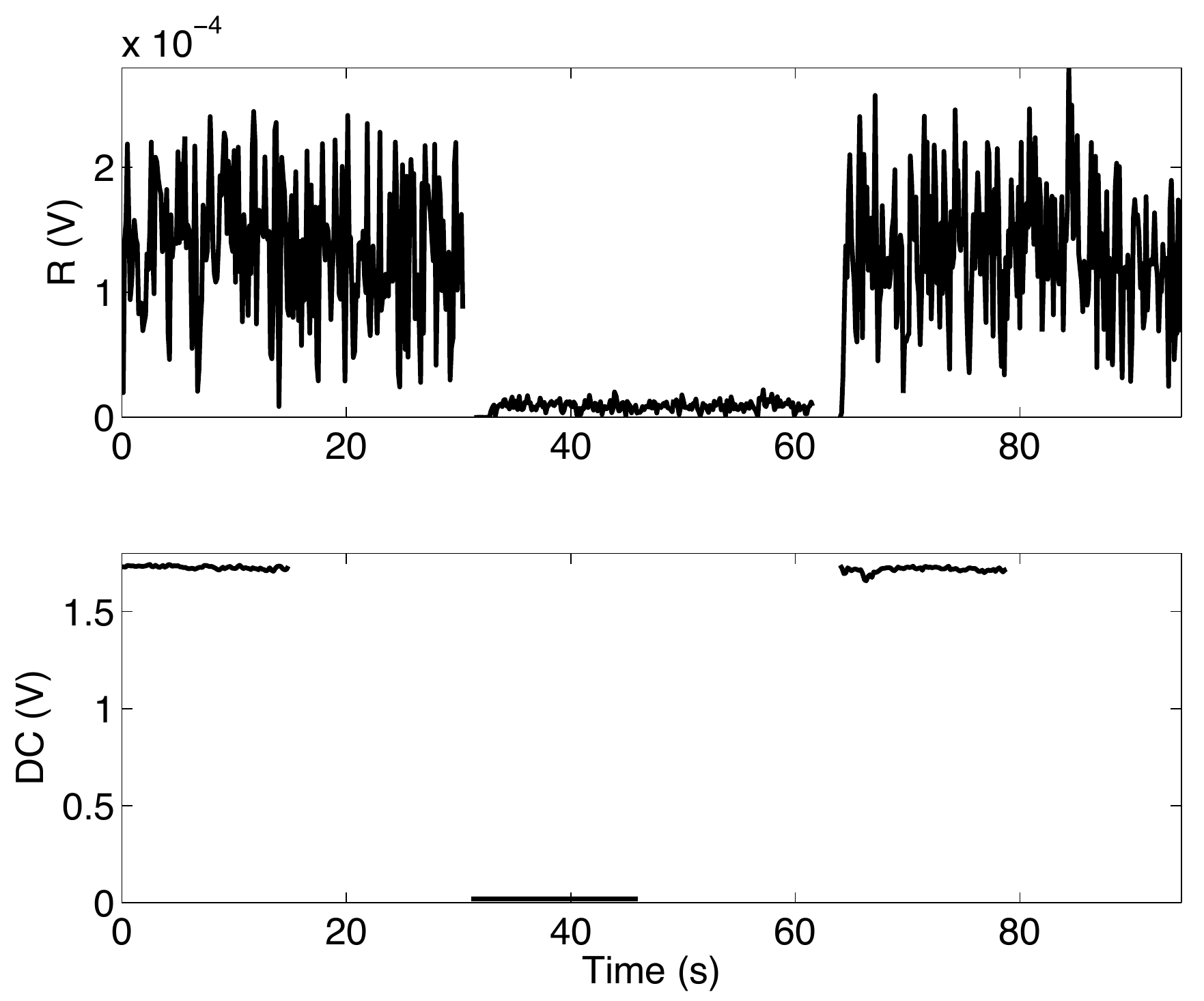}}
  \end{center}
 \caption[Typical raw AC and DC data for the unpolarized star HR 5854]{Typical raw AC and DC data for the unpolarized star HR 5854. Instantaneous values of $R$ are calculated from the observables $X$ and $Y$ via Equation \ref{2eqk}.}   
 \label{2myfigo}        
\end{figure}

\begin{figure}[p]     
 \begin{center}
  \scalebox{0.55}{\includegraphics{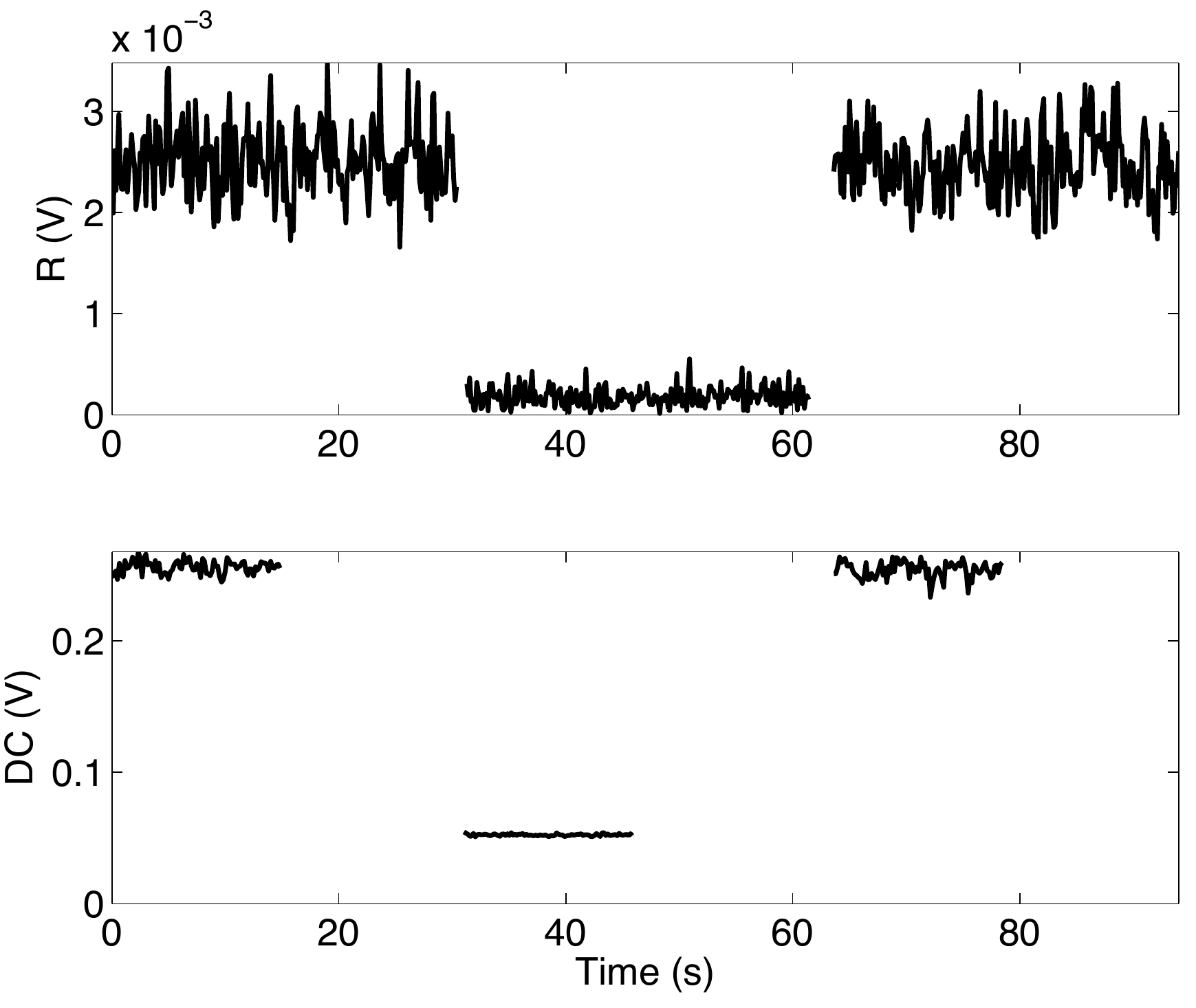}}
  \end{center}
 \caption{Typical raw data for the strongly polarized star HD 204827.}   
 \label{2myfigp}        
\end{figure}

After the target star is acquired, a measurement with 15 seconds duration is initiated on both voltmeters simultaneously with a measurement for about 30 seconds on both lock-in amplifiers. For the second half of the 30 second lock-in amplifier measurements, the voltmeters transmit data to the laptop. The secondary mirror is chopped 25 arcsec north with respect to the target star for sky subtraction, and another set of voltmeter and lock-in amplifier measurements is begun. After this has completed, the target star is returned to the field of view and another set of measurements begins. An integration ``triplet" is defined to be an on-source, 30 second AC integration and a 15 second DC integration both before and after the same measurement on a sky field. Sky levels in the optical are small compared to target star levels, so sky fields are observed with an asymmetric, 2:1 source to sky chop. See Figure \ref{2myfigo} for a typical AC and DC measurement of the unpolarized HR 5854, and see Figure \ref{2myfigp} for the strongly polarized HD 204827. The increase in AC level, and therefore in polarization, is striking between these stars.\\

Since two light beams with perpendicular polarization orientations experience the same retardance when passed through the PEM, the same polarization should be observed for the PEM at $\theta_{\rm PEM} = \pm 45^\circ$ with respect to the optical axis of the instrument. We rotate the PEM between these two positions to investigate the systematics of the PEM. The PEM is mounted to a gear driven by a stepper motor with an 8:1 step ratio, where the center of the PEM aperture is coincident with the rotation axis of the gear. Each motor step corresponds to a rotation of the PEM by $\Delta\theta_{\rm PEM} = 0 \, \fdg 1125$. For a Cassegrain ring angle of $\phi = 0^\circ$, the home position of the PEM projects its compression/extension axis northeast onto the sky (the $+U$ direction). This will be referred to as the ``PEM $+45^\circ$" position. The ``PEM $-45^\circ$" position causes this projection to be northwest on the sky (the $-U$ direction).\\

We rotate the Cassegrain ring and instrument through $\Delta\phi = 360^\circ$ to investigate rotation systematics, and PEM systematics are investigated by rotating the PEM to the $\pm 45^\circ$ positions (sections 6.2 and 6.3). The precision of the Cassegrain ring angle is $\sigma_\phi = 0.1^\circ$. A standard observing sequence begins with the Cassegrain ring at $\phi = +180^\circ$ and an integration triplet at the PEM $+45^\circ$ position followed by a triplet at the PEM $-45^\circ$ position. The ring angle is then decremented by $\Delta\phi = 45^\circ$, at which point PEM $-45^\circ$ and PEM $+45^\circ$ triplets are taken. This process occurs for each target star for Cassegrain ring angles of $+180^\circ > \phi > -180^\circ$ in $\Delta\phi = 45^\circ$ increments to sample all $\pm Q/I$ and $\pm U/I$ Stokes components. \\

The next star will see the ring angles begin at $\phi = -180^\circ$ and end at $\phi = +180^\circ$. The endpoints of $\phi = \pm180^\circ$ ensure that the ring will not ``wind up" and be forced to de-rotate during an observing sequence, wasting observing time. After each triplet, either the PEM or Cassegrain ring is rotated but not both. Each standard star is generally given eight Cassegrain ring rotations ($\Delta\phi = 360^\circ$) at two PEM positions each ($\theta_{\rm PEM} = \pm45^\circ$), and two chop integrations are taken at each PEM position. Thus, most standard stars receive about 16 minutes of AC data and about 8 minutes of DC data per night.

\section{Data Reduction}
 
Mean $X$, $Y$, and DC values for each detector are found for all on-source and sky measurements. The mean on-source values are then subtracted by the mean sky values. Assuming Stokes $Q/I$ is observed, the polarization is calculated by the following (Appendix A):

\begin{equation}
\frac{Q}{I} = \frac{\sqrt{2}}{E_{\rm{PEM}}} \frac{\sqrt{\left(X_{\rm{src}} - X_{\rm{sky}}\right)^2+\left(Y_{\rm{src}} - Y_{\rm{sky}}\right)^2}}{{\rm DC}_{\rm{src}} - {\rm DC}_{\rm{sky}}}
\label{2eqj}
\end{equation}
 \smallskip

\noindent The efficiency of the PEM, $E_{\rm PEM}$, is the strength of the AC signal based on the choice of PEM peak retardance, and it is derived in Appendix A. For POLISH, this efficiency is $E_{\rm PEM} \approx 86\%$. The sign of the final polarization of each on-source measurement is multiplied by the sign of the Stokes parameter measured. That is, the sign of the calculated Stokes parameter is calibrated by the phase from the lock-in amplifiers (Equation \ref{2eql}). \\

To calibrate absolute polarization measured with POLISH, we compare our measured polarization degree with the values of Heiles (2000) for all stars (section 8). Gain factors of $G_{\rm APD} \approx 1.56$ and $G_{\rm PMT} \approx 1.26$ are therefore multiplied to APD and PMT derived polarizations, respectively. Since the bandpasses of the red and blue enhanced APDs are nearly identical, the polarization gain factors for the APDs are not significantly different. However, the large difference in bandpass between the APDs and PMTs causes the significant difference in the polarization gain factors between the APDs and PMTs. Position angle of net polarization is not calibrated for, because our observations are generally consistent with those in the literature for strongly polarized targets. \\

Nightly mean and run-averaged $Q/I$ and $U/I$ for each source are determined by taking the weighted mean polarization of all data over the requested timescale. The weighting for each measurement is its sky-subtracted DC level multiplied by integration time. This value is proportional to the total number of detected photons and ensures that all detected photons, rather than all measurements, are weighted equally. This is important for data taken in partly cloudy conditions. The polarimetric uncertainty is the square root of the weighted variance divided by the square root of the number of measurements. It is important to note that precision calculated in this manner is only applicable to stars with no intrinsic polarimetric variability. Analyses of the variability of the observed stars, including Cygnus X-1, will be made in a future paper. \\

Expected photon shot noise, detector noise, and observed noise are derived in Appendix B as Equations \ref{2eqau} through \ref{2eqat}: 
 
  \begin{mathletters}
 \begin{equation}
 \sigma_{P \rm{shot}} = \frac{\gamma_0 \sqrt{2}}{E_{\rm{PEM}} {\rm DC}} \left\{ \frac{1}{t_{\rm{AC}}} \left[ B_{\rm AC} +\frac{1}{2} {\rm  min} \Big(B_{\rm max},B \Big) \Big( E_{\rm{PEM}} P  \Big)^2 \right] \right\} ^{1/2}
 \label{2eqau}
 \end{equation}
 \begin{equation}
 \sigma_{P \rm{detector}} = \frac{\gamma \sqrt{2}}{E_{\rm{PEM}} {\rm DC}} \left\{ \frac{1}{t_{\rm{AC}}} \left[ B_{\rm AC} +\frac{1}{2} {\rm  min} \Big(B_{\rm max},B \Big) \Big( E_{\rm{PEM}} P  \Big)^2 \right] \right\} ^{1/2}
 \end{equation}
 \begin{equation}
 \sigma_{P \rm{obs}} = \frac{\sqrt{2}}{E_{\rm{PEM}} {\rm DC}} \left\{ \frac{1}{t_{\rm{AC}}} \left[ \frac{X^2 \sigma_X^2 + Y^2 \sigma_Y^2}{X^2 + Y^2} +\frac{1}{2} \Big( E_{\rm{PEM}} P \sigma_{\rm{DC}} \Big)^2 \right] \right\} ^{1/2}
 \label{2eqat}
 \end{equation}
  \begin{equation}
  B_{\rm max} = \left(\frac{\rm{DC}}{e G T_A}\right)^{1/2}
  \end{equation}
  \end{mathletters}
 \smallskip
 
\noindent Here, $\gamma_0 \equiv 2 e G T_A {\rm DC}$, $\gamma \equiv 2 e G^{1 + x} T_A ({\rm DC} + i'_d T_A)$, $e$ is the electron charge, $B_{\rm AC} \approx 2.6$ Hz is the bandwidth of the lock-in amplifiers, and $t_{\rm AC}$ is the integration time of the lock-in amplifiers (Figures \ref{2myfigo} and \ref{2myfigp}). The values $i'_d$, $x$, $B$, $G$, and $T_A$ are the detector's output dark current, excess noise factor, bandwidth, gain, and transimpedance, which are listed in Table \ref{2tablea}. A perfect detector will have noiseless gain, $x = 0$, and dark current $i'_d = 0$. Consultation of $x$ for both detector types shows that gain from the PMTs is an order of magnitude less noisy than from the APDs, which is why PMTs are preferred over APDs for faint objects. Each of the uncertainties $\sigma_{X, Y, {\rm DC}}$ represents the sample standard deviation of $X$, $Y$, or DC of the source added in quadrature to that of the sky. The values $X$, $Y$, and DC in Equation \ref{2eqat} are sky-subtracted. \\

\begin{deluxetable}{c c c c c c}
\tabletypesize{\small}
\tablecaption{Detector Quantities}
\tablewidth{0pt}
\tablehead{
\colhead{Detector} & \colhead{$G$} & \colhead{$x$} & \colhead{$T_A$} & \colhead{$B$} & \colhead{$i'_d$} \\
 & & & (V/A) & (kHz) & (nA) }
 \startdata
 Blue APD & 300 & 0.138 & $4\times10^6$ & 90 & 3.5\\
 Red APD & 220 & 0.138 & $4\times10^6$ & 100 & 5.6\\
 PMT & $5\times10^6$ & 0.013 & $10^5$ & 200 & 0.1\\
 \enddata
\label{2tablea}
\end{deluxetable}

\section{Standard Stars and Systematic Effects}
\subsection{Telescope Polarization}
 
\begin{figure}[t]     
 \begin{center}
  \scalebox{0.65}{\includegraphics{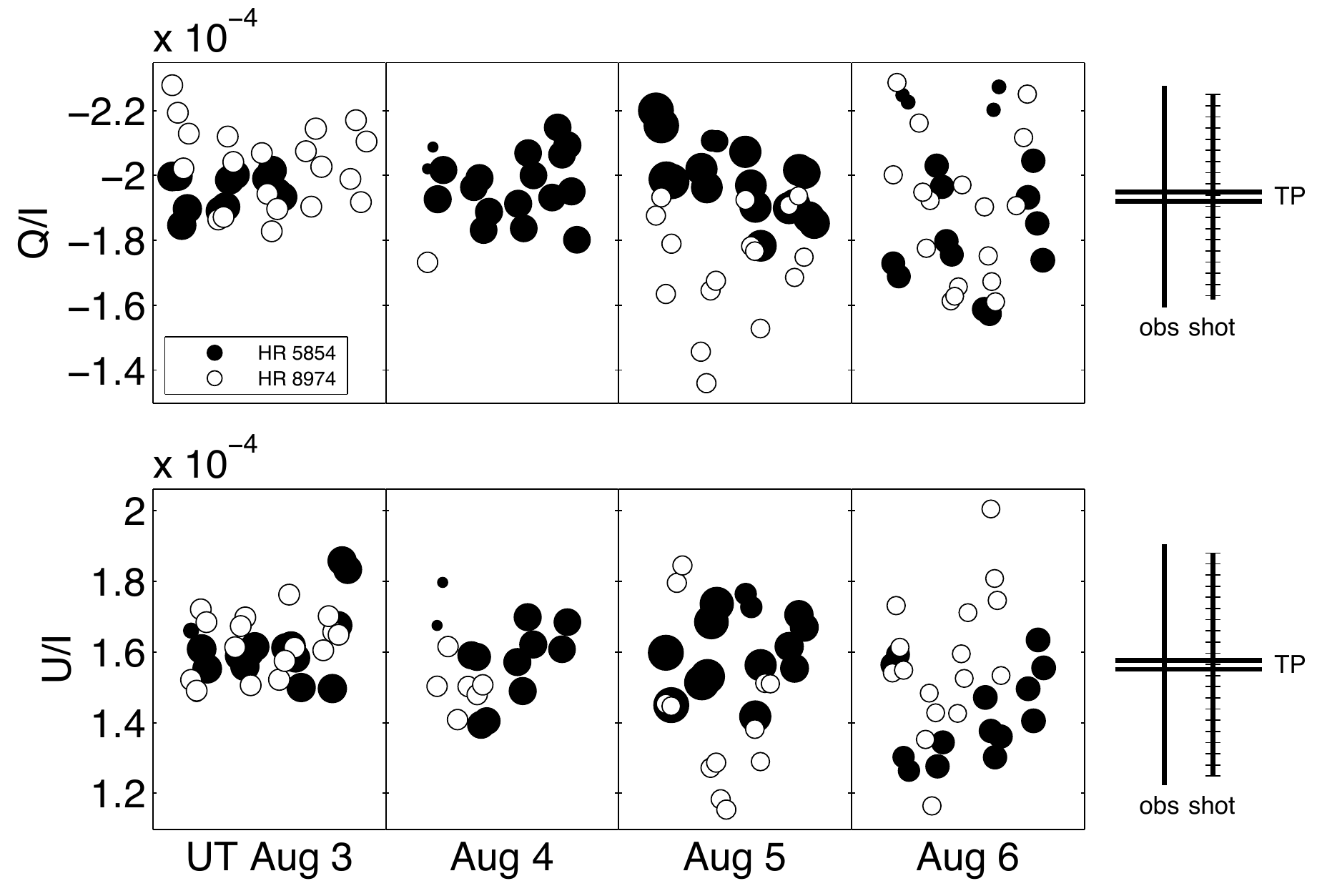}}
  \end{center}
 \caption{Raw, individual measurements of HR 5854 and HR 8974 to determine telescope polarization. The area of the data points is proportional to the number of detected photons. The vertical lines marked ``obs" indicate the median, $1\sigma$ uncertainty in polarization of the data points calculated from Equation \ref{2eqat}. The vertical lines marked ``shot" denote the expected, median photon shot noise from Equation \ref{2eqau}. The horizontal lines outside the plot boxes indicate the $\pm 1\sigma$ uncertainty in run-averaged telescope polarization for each Stokes parameter.}   
 \label{2myfigh}        
\end{figure}

From Table \ref{2tablec}, the polarizations of both HR 5854 (HD 140573) and HR 8974 (HD 222404) are close to zero, which makes them candidates for being truly unpolarized sources. Individual, shot noise limited measurements of HR 5854 and HR 8974, before subtraction of telescope polarization, are plotted in Figure \ref{2myfigh}. Individual measurements for both stars lie within error of each other, and nightly mean Stokes parameters for both stars do not generally differ significantly. The equatorial mount of the Hale 5-m inhibits traditional telescope polarization measurement, which involves allowing the field to rotate and determining the center of the $(Q,U)$ locus. However, since HLB06 performed this analysis and claim part per million polarization of HR 5854 and HR 8974, we therefore assume that these stars are indeed unpolarized. Lucas et al. (2008) report stability of HR 5854 after one year at the level of $\Delta Q,U \approx 5 \pm 2$ parts per million, indicating only weak evidence for part per million variability. \\

Nightly and run-averaged telescope polarization are calculated by the weighted mean polarization from HR 5854 and HR 8974, and they are listed in Table \ref{2tablee} as ``TP". Uncertainties in these estimates are given as the square root of the weighted variance of the individual measurements divided by the square root of the number of measurements. The light beam from the secondary mirror impinges directly on the PEM, so we assume that instrumental polarization is negligible. Indeed, the similar instrument PlanetPol has an intrinsic polarization of only a few parts per million (HLB06). \\

\begin{deluxetable}{c c c c c c}
\tabletypesize{\small}
\tablecaption{Telescope Polarization from Unpolarized Standard Stars}
\tablewidth{0pt}
\tablehead{
\colhead{UT Date} & \colhead{Target} & \colhead{$Q/I$} & \colhead{$U/I$} & \colhead{$P$} & \colhead{$\Theta$} \\
& & $\left(\times10^{-6}\right)$ & $\left(\times10^{-6}\right)$ & $\left(\times10^{-6}\right)$ & $\left(^\circ\right)$ }
\startdata
2007 Aug 3 & HR 5854 & $-$195.1(1.5) & +161.9(2.5) & 253.5(2.0) & 70.16(25) \\
2007 Aug 4 & $\cdots$ & $-$196.6(2.3) & +156.9(2.9) & 251.6(2.5) & 70.70(31) \\
2007 Aug 5 & $\cdots$ & $-$198.8(2.5) & +159.4(2.8) & 254.8(2.6) & 70.65(30) \\
2007 Aug 6 & $\cdots$ & $-$185.2(4.9) & +142.8(3.2) & 233.9(4.3) & 71.17(48) \\
Overall & $\cdots$ & $-$195.3(1.5) & +156.5(1.7) & 250.2(1.6) & 70.64(19) \\
\hline
2007 Aug 3 & HR 8974 & $-$202.9(2.7) & +163.7(1.9) & 260.8(2.4) & 70.55(25) \\
2007 Aug 4 & $\cdots$ & $-$173(31) & +149.7(2.9) & 229(24) & 69.6(2.6) \\
2007 Aug 5 & $\cdots$ & $-$170.0(5.8) & +140.2(5.0) & 220.3(5.5) & 70.24(69) \\
2007 Aug 6 & $\cdots$ & $-$189.1(4.7) & +170.2(4.5) & 254.4(4.6) & 69.00(52) \\
Overall & $\cdots$ & $-$189.3(3.1) & +156.2(2.4) & 245.5(2.9) & 70.24(32) \\
\hline
2007 Aug 3 & TP & $-$198.8(1.8) & +162.6(1.7) & 256.8(1.8) & 70.37(19) \\
2007 Aug 4 & $\cdots$ & $-$195.8(2.4) & +155.7(2.4) & 250.1(2.4) & 70.75(27) \\
2007 Aug 5 & $\cdots$ & $-$191.6(3.4) & +154.6(2.9) & 246.2(3.2) & 70.55(36) \\
2007 Aug 6 & $\cdots$ & $-$186.6(3.5) & +149.0(3.2) & 238.9(3.4) & 71.70(40) \\
Overall & $\cdots$ & $-$193.5(1.5) & +156.4(1.3) & 248.9(1.4) & 70.53(16) \\
\enddata
\label{2tablee}
\tablecomments{These polarization values are the mean, weighted by DC level, of measurements taken by both the red and blue enhanced APDs. Therefore, they are measured in the wavelength range of about 400 nm to roughly 850 nm. Uncertainty in position angle is purely statistical, as position angle is not calibrated absolutely.}
\end{deluxetable}

The cause of the large telescope polarization of the Hale 5-m, $\rm{TP} \approx 2.5 \times 10^{-4}$, may be due to inhomogeneities in the coating of the primary and/or secondary mirrors. Net stellar polarizations are shown in Tables \ref{2tableh} and \ref{2tablei} after nightly telescope polarization is subtracted. Weakly polarized stars ($P < 0.1\%$) are given in Table \ref{2tableh}, while strongly polarized stars ($P > 0.1\%$) are listed in Table \ref{2tablei}. Since HR 5854 and HR 8974 are effectively unpolarized, uncertainty in polarimetric position angle $\Theta$ is so large as to preclude meaningful estimates on $\Theta$. \\

\begin{deluxetable}{c c c c c c}
\tabletypesize{\footnotesize}
\tablecaption{Weakly Polarized Standard Stars}
\tablewidth{0pt}
\tablehead{
\colhead{UT Date} & \colhead{Star} & \colhead{$Q/I$}  & \colhead{$U/I$}  & \colhead{$P$}  & \colhead{$\Theta$} \\
& & $\left(\times10^{-6}\right)$ & $\left(\times10^{-6}\right)$ & $\left(\times10^{-6}\right)$ & $\left(^\circ\right)$ }
\startdata
2007 Aug 3 & HR 5854 & +3.7(2.4) & $-$0.8(3.0) & 3.8(2.4) & $-$ \\
2007 Aug 4 & $\cdots$ & $-$0.7(3.3) & +1.9(3.8) & 2.0(3.8) & $-$ \\
2007 Aug 5 & $\cdots$ & $-$7.7(4.2) & +4.2(4.0) & 8.8(4.1) & $-$ \\
2007 Aug 6 & $\cdots$ & +1.7(6.0) & $-$6.8(4.5) & 7.1(4.6) & $-$ \\
Overall & $\cdots$ & $-$2.0(2.2) & +0.3(2.2) & 2.0(2.2) & $-$ \\
\hline
2007 Aug 3 & HR 8974 & $-$3.2(3.3) & +1.3(2.5) & 3.5(3.1) & $-$ \\
2007 Aug 4 & $\cdots$ & +22.6(2.4) & $-$6.1(3.8) & 23.3(2.5) & $-$ \\
2007 Aug 5 & $\cdots$ & +21.6(6.7) & $-$14.4(5.7) & 25.9(6.4) & $-$ \\
2007 Aug 6 & $\cdots$ & $-$2.5(5.9) & +21.6(5.5) & 21.7(5.5) & $-$ \\
Overall & $\cdots$ & +4.7(3.5) & $-$0.1(2.8) & 4.7(3.5) & $-$ \\
\hline
2007 Aug 4 & HD 9270 & $-$33.9(5.1) & $-$76.0(4.0) & 83.1(4.3) & 123.0(1.7) \\
2007 Aug 5 & $\cdots$ & $-$37.6(7.1) & $-$82.2(4.2) & 90.4(4.8) & 122.7(2.1) \\
2007 Aug 6 & $\cdots$ & $-$43.0(8.4) & $-$81.6(6.9) & 92.1(7.3) & 121.2(2.6) \\
Overall & $\cdots$ & $-$35.0(3.5) & $-$81.8(2.7) & 88.9(2.8) & 123.4(1.1) \\
\hline
2007 Aug 4 & $\gamma$ Oph & $-$86.5(5.6) & +134.9(4.3) & 160.3(4.7) & 61.33(94) \\
2007 Aug 5 & $\cdots$ & $-$81.9(4.0) & +136.9(4.9) & 159.5(4.7) & 60.45(76) \\
2007 Aug 6 & $\cdots$ & $-$65.8(7.7) & +118.5(7.1) & 135.5(7.2) & 59.5(1.6) \\
Overall & $\cdots$ & $-$77.6(3.5) & +128.4(3.4) & 150.1(3.4) & 60.56(65) \\
\hline
2007 Aug 3 & HD 212311 & +330(63) & $-$4(60) & 330(63) & 179.6(5.2) \\
2007 Aug 4 & $\cdots$ & +260(48) & +15(60) & 261(48) & 1.6(6.6) \\
2007 Aug 5 & $\cdots$ & +300(52) & $-$61(68) & 306(52) & 174.5(6.4) \\
2007 Aug 6 & $\cdots$ & +178(58) & $-$136(68) & 226(65) & 161.1(8.9) \\
Overall & $\cdots$ & +273(29) & $-$33(34) & 275(29) & 176.5(3.6) \\
\hline
2007 Aug 5 & Algenib & $-$526(10) & $-$528(11) & 745(11) & 112.54(41) \\
2007 Aug 6 & $\cdots$ & $-$579.2(7.3) & $-$512.7(7.2) & 773.5(7.2) & 110.76(27) \\
Overall & $\cdots$ & $-$557.1(6.8) & $-$521.8(5.8) & 763.3(6.4) & 111.56(24) \\
\enddata
\label{2tableh}
\tablecomments{The polarization values for all stars except HD 212311 are the mean, weighted by DC level, of measurements taken by both the red and blue enhanced APDs. Therefore, they are measured in the wavelength range of about 400 nm to roughly 850 nm. The polarization values for HD 212311 were taken by the PMTs. Therefore, they are measured in the wavelength range of about 400 nm to 675 nm. Telescope polarization has been subtracted from all values (section 6.1). Uncertainty in position angle is purely statistical, as position angle is not calibrated absolutely.}
\end{deluxetable}

\begin{deluxetable}{c c c c c c}
\tabletypesize{\small}
\tablecaption{Strongly Polarized Standard Stars}
\tablewidth{0pt}
\tablehead{
\colhead{UT Date} & \colhead{Star} & \colhead{$Q/I$}  & \colhead{$U/I$}  & \colhead{$P$}  & \colhead{$\Theta$} \\
& & $\left(\%\right)$ & $\left(\%\right)$ & $\left(\%\right)$ & $\left(^\circ\right)$ }
\startdata
2007 Aug 4 & u Her & +0.1309(13) & $-$0.03803(81) & 0.1363(13) & 171.90(18) \\
\hline
2007 Aug 5 & HD 157999 & $-$0.8763(12) & +0.1531(20) & 0.8895(12) & 85.045(65) \\
2007 Aug 6 & $\cdots$ & $-$0.86422(93) & +0.1492(24) & 0.8770(10) & 85.103(75) \\
Overall & $\cdots$ & $-$0.8699(13) & +0.1509(16) & 0.8828(13) & 85.079(51) \\
\hline
2007 Aug 3 & HD 187929 & $-$1.5949(10) & $-$0.1610(56) & 1.6030(12) & 92.883(99) \\
2007 Aug 5 & $\cdots$ & $-$1.5551(27) & $-$0.1758(41) & 1.5651(28) & 93.224(75) \\
2007 Aug 6 & $\cdots$ & $-$1.5262(27) & $-$0.1727(40) & 1.5360(28) & 93.229(76) \\
Overall & $\cdots$ & $-$1.5784(37) & $-$0.1659(35) & 1.5872(37) & 93.001(65) \\
\hline
2007 Aug 4 & HD 7927 & $-$3.0658(40) & $-$0.2510(93) & 3.0761(40) & 92.342(87) \\
\hline
2007 Aug 3 & HD 147084 & $-$ & +3.3972(47) & $-$ & $-$ \\
2007 Aug 4 & $\cdots$ & +1.658(14) & +3.4386(25) & 3.8171(65) & 32.134(95) \\
Overall & $\cdots$ & $-$ & +3.4050(56) & $-$ & $-$ \\
\hline
2007 Aug 4 & HD 154445 & $-$3.8076(67) & $-$0.0405(12) & 3.8078(67) & 90.3047(86) \\
\hline
2007 Aug 3 & HD 204827 & $-$2.6413(82) & +4.733(89) & 5.4202(89) & 59.582(46) \\
2007 Aug 4 & $\cdots$ & $-$ & +4.819(12) & $-$ & $-$ \\
2007 Aug 5 & $\cdots$ & $-$2.638(11) & +4.742(10) & 5.427(10) & 59.542(56) \\
2007 Aug 6 & $\cdots$ & $-$2.6638(89) & +4.7970(95) & 5.4869(95) & 59.521(47) \\
Overall & $\cdots$ & $-$2.6463(56) & +4.7575(68) & 5.4440(66) & 59.542(31) \\
\enddata
\label{2tablei}
\tablecomments{The polarization values for all stars except HD 204827 are measured in the wavelength range of about 400 nm to roughly 850 nm. The polarization values for HD 204827 are measured in the wavelength range of about 400 nm to 675 nm. Telescope polarization has been subtracted from all values (section 6.1). Uncertainty in position angle is purely statistical, as position angle is not calibrated absolutely.}
\end{deluxetable}
 
 Even though the bandpasses differ between the APDs and PMTs, we do not determine telescope polarization with the PMTs for many reasons. First, it is difficult to identify unpolarized stars with $V > 7$. Second, differences in telescope polarization derived from APD and PMT observations will only be detected after long PMT observations. Third, we aim to detect small scale changes in polarization of target stars, so constant offsets in telescope polarization between APD and PMT observations is not our goal. Therefore, we choose to quickly measure telescope polarization at the part per million level using bright stars and APDs to minimize overhead due to calibration. \\

\subsection{PEM Systematics}

The PEM is rotated to positions of $\theta_{\rm PEM} = \pm45^\circ$ with respect to the Wollaston axis, and the Cassegrain ring is rotated through $\Delta\phi = 360^\circ$. This gives independent measures of the PEM and ring rotation systematics. While PlanetPol can be rotated to positions of $\pm45^\circ$ with respect to their PEM, this $90^\circ$ rotation of the instrument also causes the Stokes parameter of opposite sign to be observed. Thus, PEM and instrument rotation systematics are coupled for PlanetPol, while they can be independently measured for POLISH. In addition, PlanetPol can only rotate through $\Delta\phi = 135^\circ$, but the $\Delta\phi = 360^\circ$ rotation of POLISH enables more thorough measurement of the instrument rotation systematics.\\

To investigate the PEM systematics, we subtract the offset due to ring rotation systematics. We first find the weighted mean polarization of each Stokes parameter for each star separately and at each of the $\theta_{\rm PEM} = \pm45^\circ$ positions. In increments of $\Delta\phi = 90^\circ$, we average ring angles $\phi = 0^\circ$ to $\phi = 270^\circ$ for the $Q/I$ parameter and $\phi = 45^\circ$ to $\phi = 315^\circ$ for the $U/I$ parameter. Therefore, the mean polarization at each PEM position contains the same offset due to ring rotation systematics. The sign of the polarization taken at the $\theta_{\rm PEM} = -45^\circ$ position is reversed, and the unweighted mean is taken across both Stokes parameters and both PEM positions. The unweighted mean is employed so neither Stokes parameter nor PEM position dominates. This value is the PEM offset, given by $S_{\rm PEM}$ (Equation \ref{2eqa}). The uncertainty in this offset is half the difference between the results for Stokes $Q/I$ and $U/I$ (Equation \ref{2eqb}). Here, the index $i$ represents the PEM position, where $i = 0$ indicates $\theta_{\rm PEM} = +45^\circ$ and $i = 1$ indicates $\theta_{\rm PEM} = -45^\circ$. This process is duplicated for each detector and star separately. \\

 \begin{mathletters}
\begin{equation}
S_{\rm{PEM,}\phi} = \frac{1}{4} \sum_{i=0}^{1}\left(-1^i\right)\overline{Q}_{i}+\left(-1^i\right)\overline{U}_{i}
\label{2eqa}
\end{equation}
\begin{equation}
\sigma_{\rm{PEM,}\phi}=  \frac{1}{4} \left|\sum_{i=0}^{1}\left(-1^i\right)\overline{Q}_{i}-\left(-1^i\right)\overline{U}_{i}\right|
\label{2eqb}
\end{equation}
 \end{mathletters}
 \smallskip

\subsection{Positive/Negative Stokes Systematics}

To investigate the systematics when rotating the Cassegrain ring by $90^\circ$, i.e. the differences between $\pm Q$ as well as between $\pm U$, we subtract the offset due to PEM systematics. While this value has been calculated above, we prefer to combine the data in such a way as to cause it to cancel. We find the weighted mean value of each Stokes parameter separately using both $\theta_{\rm PEM} = \pm45^\circ$ positions. We average ring angles $\phi = 0^\circ$ and $\phi = 180^\circ$ for the $+Q/I$ parameter, $\phi = 90^\circ$ and $\phi = 270^\circ$ for $-Q/I$, $\phi = 45^\circ$ and $\phi = 225^\circ$ for $+U/I$, and $\phi = 135^\circ$ and $\phi = 315^\circ$ for $-U/I$. We then reverse the signs of polarization for the negative Stokes parameters. Taking the unweighted mean for $Q/I$ and $U/I$ separately, we find the offsets for both Stokes parameters, $S_\phi$ (Equation \ref{2eqa}). The uncertainty is half the difference between the offsets for the positive and negative Stokes parameters (Equation \ref{2eqb}). Here, the index $i$ represents the sign of the measured Stokes parameter, where $i = 0$ indicates $+Q,+U$ and $i = 1$ indicates $-Q,-U$. \\

Both systematics are calculated for each detector combination. A weighted mean is taken across all stars to determine the mean systematics of the instrument/telescope system. Weighting is given as the inverse square of the individual uncertainties, as opposed to weighting by detected photons. This is because mean values are taken across all stars as opposed to across all measurements of a particular star. The mean systematic across all stars when rotating the PEM by $\pm 90^\circ$ is $S_{\rm PEM} = (+1.6 \pm 4.7) \times 10^{-6}$, while the mean systematic between positive and negative Stokes parameters is $S_\phi = (-3.0 \pm 2.8) \times 10^{-6}$. The quadrature addition of $S_{\rm PEM}$ and $S_\phi$ divided by stellar polarization has a mean value of $S / P = 0.61 \pm 0.21\%$. Therefore, systematic effects are consistent with zero and are only important at less than $1\%$ of the measured polarization. \\

\section{Run-Averaged Precision}

\begin{figure}[t]     
 \begin{center}
  \scalebox{0.85}{\includegraphics{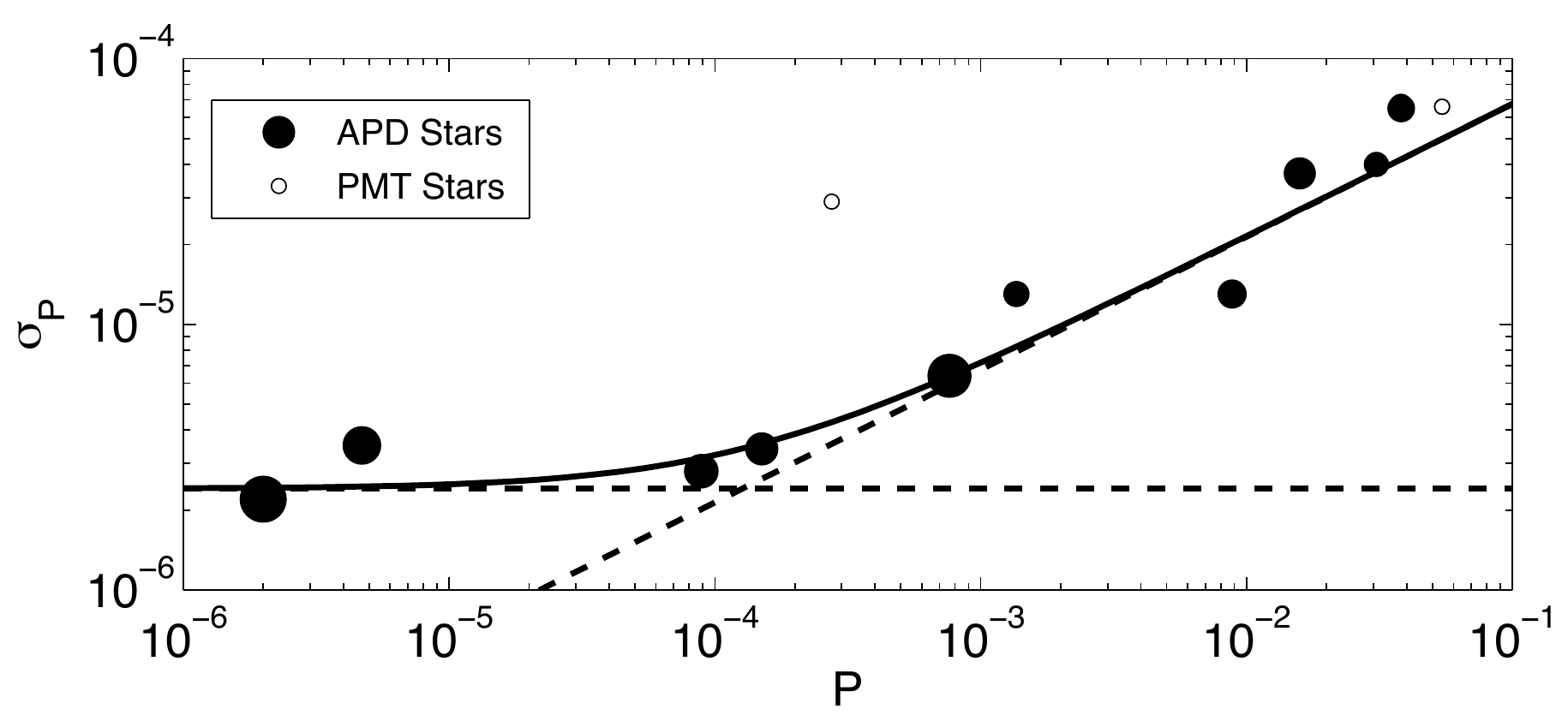}}
  \end{center}
 \caption[Run-averaged precision as a function of stellar polarization]{Run-averaged precision as a function of stellar polarization. Photon shot noise considerations predict precision proportional to the square root of polarization, which is observed. The solid line is a fit to the data with power law slope 1/2 plus the quadrature addition of an instrumental noise floor. The dashed lines are the $P^{1/2}$ and noise floor terms. The diameters of the data points are inversely proportional to stellar magnitude.}   
 \label{2myfign}        
\end{figure}

The flux of net polarized light from a star is proportional to the polarization, $P$. Since the photon shot noise on this quantity scales as $P^{1/2}$, one would expect that the run-averaged, polarimetric precision attainable on stars of similar brightness would also be proportional to $P^{1/2}$. In addition, we expect the instrument to have a noise floor that becomes noticeable for unpolarized stars. As can be seen in Figure \ref{2myfign}, we find good agreement by fitting the data from the stars observed with APDs to the model

\begin{equation}
\hat{\sigma}_P = \left[\left(\frac{P^{1/2}}{a}\right)^2 + \sigma^2_{0}\right]^{1/2}
\end{equation}
\smallskip

\noindent Here, $a$ is a scaling factor and $\sigma_{0} = 2.4$ parts per million is the noise floor of the instrument. The fitting was performed using a least-squares approach. However, since the data span five orders of magnitude in polarization, the residuals to be minimized are given by

\begin{equation}
{\rm SSE} = \sum_{i}\left(\frac{ \sigma_{i} - \hat{\sigma}}{\sigma_{i}} \right)^2
\end{equation}
\smallskip

The stars observed with APDs are all roughly the same visual magnitude. However, the precision achieved on the weakly polarized HD 212311, observed with PMTs, is about an order of magnitude worse than for the stars observed with APDs. This is expected, since HD 212311 is roughly 5 magnitudes fainter than the APD stars (Table \ref{2tablec}). Thus, the scaling factor $a$ determined for the APD stars will be an order of magnitude different from the scaling factor for the PMT stars. This is why the PMT stars were excluded from the above fit. However, precision on the strongly polarized HD 204827 is surprisingly consistent with the slope for the bright stars. The four stars with polarization $P > 3\%$, HD 7927, HD 147084, HD 154445, and HD 204827, are known to be variable (Dolan \& Tapia 1986; Bastien et al. 1988, 2007). Thus, differences in $\sigma_P$ and the model fit are likely to be related to stellar variability. As stated in section 5, it should be noted that run-averaged precision assumes nonvariable polarization of the target. The validity of this assumption will be investigated in a future paper. \\

\section{Comparison to Literature}
 
\begin{figure}[p]     
\begin{center}
\scalebox{0.9}{\includegraphics{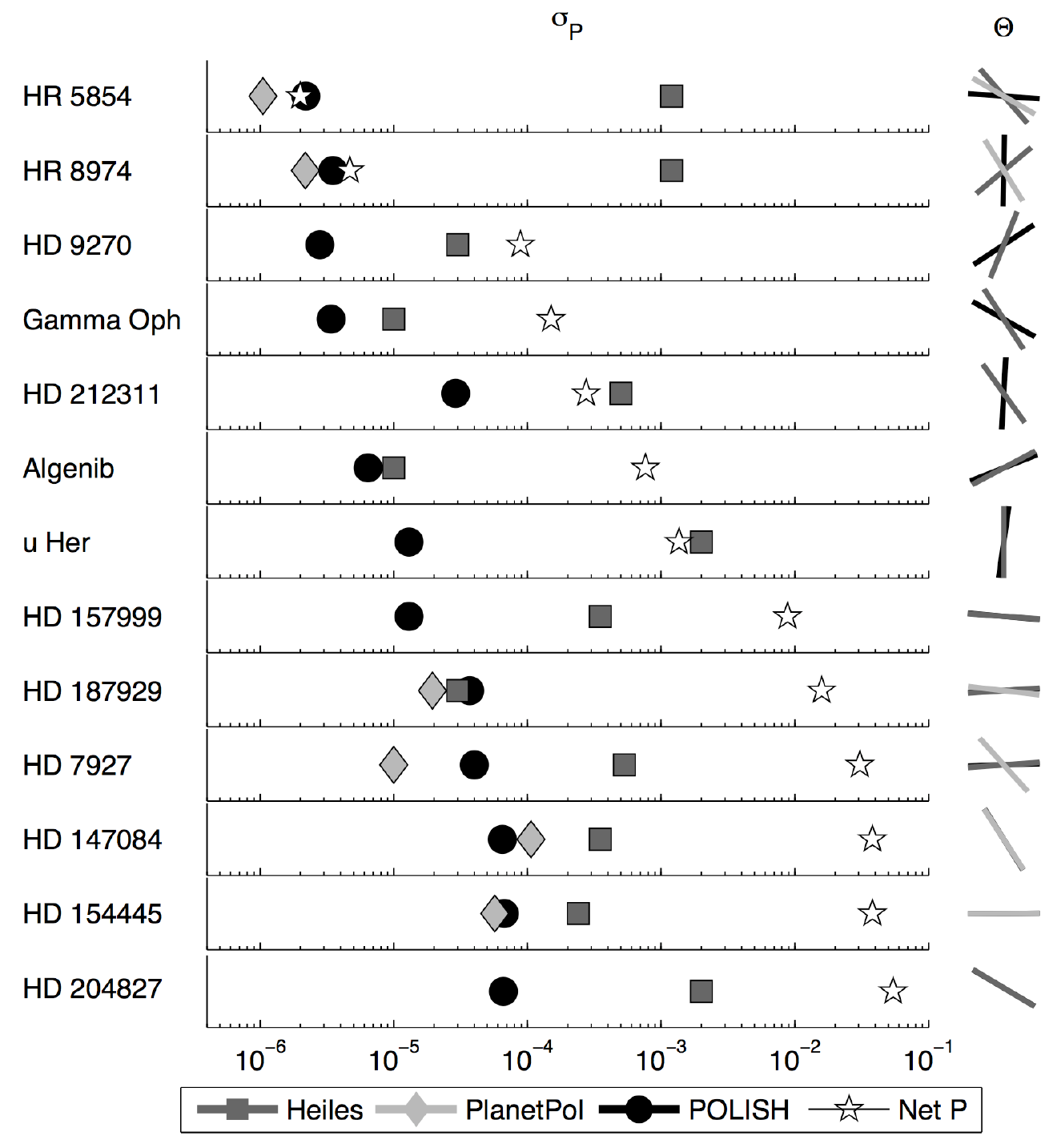}}
\end{center}
\caption[Precision achieved on standard stars compared to the literature]{Precision achieved on standard stars compared with HLB06 and Heiles (2000). It must be stressed that POLISH observations are unfiltered, HLB06 observations were taken in the wavelength range of 625 to 950 nm, and Heiles (2000) data are for $V$ band. Stars are listed from top to bottom in order of increasing net polarization according to our measurements. Note that HD 212311 and HD 204827 were observed with PMTs (about 400 to 675 nm bandpass), while the rest of the sample was observed with APDs (about 400 to 850 nm bandpass).}   
\label{2myfige}        
\end{figure}

We compare our results to the agglomerated polarization catalog of Heiles (2000) and to HLB06 in Figure \ref{2myfige}. The degree of polarization measured by POLISH is plotted as open stars, and stellar polarization increases toward the bottom of the plot. Precision in the degree of polarization from this work is plotted as filled black circles, precision values computed from HLB06 are light grey diamonds, and Heiles (2000) precision values are dark grey squares. Uncertainty in degree of polarization is listed nightly for strongly polarized stars in Table 3 of HLB06 (HD 7927, HD 147084, HD 154445, and HD 187929). To determine run-averaged uncertainty for these stars, we first convert their degree, position angle, and uncertainties to $Q/I$, $U/I$, and associated uncertainties. We then perform a weighted mean for each Stokes parameter separately, where the weights are the inverse square of the nightly uncertainties in those parameters. To determine polarization for each star, Heiles (2000) take the weighted mean polarization from different authors. The weights are the inverse square of the uncertainties from each author. Uncertainty in stellar polarization in Heiles (2000) is the square root of the sum of squares of residuals between each author's polarization and the Heiles (2000) mean polarization. \\

The rightmost column of the figure shows the position angle of net polarization, where north is at the top and east is at the left of the plots. Orientations of the black lines indicate position angles measured with POLISH, HLB06 position angles are light grey lines, and Heiles (2000) position angles are dark grey lines. Agreement between the data sets for stars with low polarization is of course poor, because position angle of net polarization is meaningless for these stars. As stellar polarization increases, however, agreement in position angle also increases. Since agreement between our measurements and the literature regarding degree of polarization is determined by choice of polarimetric gain factor (section 5), accuracy in our observations is assessed by agreement in position angle of polarization. \\

The unpolarized standard stars observed in order to determine telescope polarization, HR 5854 and HR 8974, have run-averaged polarimetric precision of order one part per million. This is our precision goal for bright, unpolarized stars, and it is comparable to the precision achieved by PlanetPol. However, we have improved the precision on these stars by nearly three orders of magnitude with respect to the Heiles (2000) catalog. It should be noted that HR 8974 is known to harbor an extrasolar planet with a minimum mass of $1.60 \pm 0.13$ Jupiter masses, a period of $902.9 \pm 3.5$ days, and a semimajor axis of $2.044 \pm 0.057$ AU (Neuh\"{a}user et al. 2006). However, we expect the amplitude of the planetary polarimetric signal to be of order $10^{-8}$ or less and consequently undetectable.\\
 
 We have improved the polarimetric precision achieved on HD 9270 by an order of magnitude with respect to Heiles (2000). Precision on $\gamma$ Oph and Algenib (HD 886), however, is only slightly better than tabulated in Heiles (2000). It is expected that more observations of these stars will improve this precision. There is an order of magnitude improvement in precision on HD 212311 with respect to Heiles (2000). The precision achieved on u Her has been improved by two orders of magnitude from Heiles (2000). \\
 
 Precision on HD 157999 has been improved by an order of magnitude, while precision on HD 187929 is comparable between our measurements, those from PlanetPol, and those from Heiles (2000). This may indicate intrinsic variability of the source. Precision on HD 7927 from POLISH is of the same magnitude as that from PlanetPol, and we improve the precision with respect to Heiles (2000) by an order of magnitude. We assume that the discrepancy between the HLB06 position angle ($42.18 \pm 0.01^\circ$) and our measurements for HD 7927 is simply a typographical error in their paper. We expect that their intended value is $92\, \fdg 18$, which is close to our value.\\
 
 Precision on HD 147084 is comparable between our measurements and those from PlanetPol. We improve on the Heiles (2000) results by almost an order of magnitude. Precision on HD 154445 between our measurements, PlanetPol, and Heiles (2000) lie within a factor of $\approx 3$ of each other, which may suggest intrinsic variability of the source. We only have one night of data on this star, so this possibility will not be investigated in our forthcoming paper. Finally, we improve the precision on HD 204827 by over an order of magnitude with respect to Heiles (2000).\\
 
\section{Discussion}
 
 We have commissioned a high precision, integrated light polarimeter in order to detect variability in the optical, linear polarization of high mass X-ray binaries. This variability should be indicative of system inclination, and high precision monitoring is hoped to constrain the mass of the black holes in these systems. While results from observations of Cygnus X-1 will be forthcoming, we observe polarization of standard stars to high precision. Individual measurements of most stars are photon shot noise limited (Figure \ref{2myfigh}). Nightly mean and run-averaged polarizations of most stars are observed at precision levels comparable to PlanetPol, a similar instrument mounted on the William Herschel Telescope. \\
 
 Precision achieved on unpolarized stars is of order one part per million, which is up to three orders of magnitude more precise than the combined polarimetric catalog of Heiles (2000). We find night-to-night precision of three to ten parts per million on bright, weakly polarized standard stars $\left(10^{-4} < P < 10^{-3}\right)$. This precision increases as $\sigma_P \propto P^{1/2}$ and is of order one part in $10^5$ for stars with $1\% < P < 10\%$. Precision on strongly polarized stars is improved by up to an order of magnitude with respect to Heiles (2000). Thus, precision achieved in a night-to-night sense scales as expected from photon shot noise statistics. Systematic effects reveal themselves at the level of less than $1\%$ of the measured polarization for polarized stars. The night-to-night noise floor of the instrument appears to be $1\%$ of the telescope polarization, or about two parts per million. This noise floor is comparable to that of PlanetPol, which is about one part per million (Lucas et al. 2008). The large improvement in polarimetric precision arises from the combination of large telescope aperture, a high-quality polarization modulator, and high frequency modulation. \\
 
 In addition to observations of high mass X-ray binaries, the high precision of POLISH is well suited to polarimetric observations of extrasolar planet host stars, debris disks, and evolved stars. While detection of scattered light from hot Jupiters requires higher precision than currently attainable, knowledge of host star baseline variability is necessary. In addition, observation of planetary transits in polarized light gives information on transit geometry and the scattering properties of the host star, which are poorly known. Polarimetric observations of debris disks may constrain the size distribution of dust grains as well as the scattering properties of the disk. Evolved stars represent the intermediate stage between roughly spherical stars and strikingly asymmetric planetary nebulae. The growth of asymmetry in these systems is poorly understood, and high precision polarimetry may make significant contributions to this field. Indeed, inexpensive, high precision polarimeters are poised to impart their mark in many fields of astronomy. \\
 
 \acknowledgements
 
 SJW would like to thank Shrinivas Kulkarni, Michael Ireland, and Ian McEwan for valuable discussions on instrumentation, polarimetry, and signal processing. In addition, the help of the Palomar Observatory staff, especially Bruce Baker, Karl Dunscombe, John Henning, Greg van Idsinga, and Steve Kunsman, has been essential. We would like to thank the anonymous referee for helpful comments on this manuscript. We would like to acknowledge support from the Moore Foundation. This research has made use of the SIMBAD database, operated at CDS, Strasbourg, France. \\
 
 {\it Facilities:} \facility{Hale}.
  
  \appendix
  
  \section{Mueller Matrix for POLISH}

The Mueller matrix of POLISH is given by
 
 \begin{equation}
 M_{\rm{POLISH}} = T_{-\phi} \times M_{\rm D} \times M_{\rm L} \times M_{\rm W} \times M_{\rm B} \times T_{-\theta} \times M_{\rm PEM} \times T_{\theta} \times T_{\phi} \times M_{\rm T}
 \label{2eqp}
 \end{equation}
\smallskip

\noindent and observed polarization is related to incident polarization according to
 
 \begin{equation}
 \left( \begin{array}{c} I_{\rm obs} \\ Q_{\rm obs} \\ U_{\rm obs} \\ V_{\rm obs} \end{array} \right) = M_{\rm{POLISH}} \left( \begin{array}{c} I \\ Q \\ U \\ V \end{array} \right)
 \label{2eqq}
 \end{equation}
\smallskip
 
 \noindent Starting from the left hand side of Equation \ref{2eqp}, the matrices are the rotation matrix for Cassegrain ring angle $\phi$, the Mueller matrices for the detector window, field lenses, Wollaston prism, and beamsplitter, the rotation matrix for PEM angle $\theta \equiv \theta_{\rm PEM}$, the Mueller matrix for the PEM, and the Mueller matrix for the telescope. Subtraction of telescope polarization is necessary to calibrate for $M_{\rm T}$. \\

Since the PEM and Wollaston prism convert the polarization of the incident light into intensity modulation, the polarization state of light past the Wollaston is not our concern. The $90^\circ$ reflection of light off the beamsplitter is stable during observations, so any polarization imparted to the light by the beamsplitter is just a constant offset to the polarization. Therefore, we only consider the throughput of the detector window, field lenses, and beamsplitter, as opposed to their polarizing properties. We denote the throughput of the instrument as $E$, and it is given in section 3. The POLISH Mueller matrix becomes

\begin{eqnarray}
M_{\rm{POLISH}} = & E\left(\begin{array}{cccc}1 & 0 & 0 & 0\\0 & \cos{2\phi} & -\sin{2\phi} & 0\\0 & \sin{2\phi} & \cos{2\phi} & 0\\0 & 0 & 0 & 1\end{array} \right) \! \! \! \! \! & \times \left(\begin{array}{cccc}0.5 & \pm0.5 & 0 & 0 \\\pm0.5 & 0.5 & 0 & 0 \\0 & 0 & 0 & 0 \\0 & 0 & 0 & 0\end{array}\right) \label{2equ} \\
& \times \left(\begin{array}{cccc}1 & 0 & 0 & 0 \\0 & \cos{2\theta} & -\sin{2\theta} & 0 \\0 & \sin{2\theta} & \cos{2\theta} & 0 \\0 & 0 & 0 & 1\end{array}\right) \! \! \! \! \!  & \times\left(\begin{array}{cccc}1 & 0 & 0 & 0 \\0 & 1 & 0 & 0 \\0 & 0 & \cos\beta & \sin\beta \\0 & 0 & -\sin\beta & \cos\beta\end{array}\right) \nonumber \\
& \times\left(\begin{array}{cccc}1 & 0 & 0 & 0 \\0 & \cos{2\theta} & \sin{2\theta} & 0 \\0 & -\sin{2\theta} & \cos{2\theta} & 0 \\0 & 0 & 0 & 1\end{array}\right) \! \! \! \! \! & \times \left(\begin{array}{cccc}1 & 0 & 0 & 0\\0 & \cos{2\phi} & \sin{2\phi} & 0\\0 & -\sin{2\phi} & \cos{2\phi} & 0\\0 & 0 & 0 & 1\end{array} \right) \nonumber
\end{eqnarray}
 \smallskip
 
 \noindent The top sign in the Wollaston matrix (+) indicates the left beam which reaches detector 2, and the bottom sign ($-$) represents the right beam which reaches detector 1. \\
 
 The instantaneous retardance of the PEM is given by $\beta= \beta_0\sin\omega t$, and expanding the retardance in terms of Bessel functions gives

 \begin{mathletters}
\begin{equation}
\sin \Big(\beta_0\sin\omega t \Big)=2\sum_{n=0}^{\infty}J_{2n+1}\left(\beta_0\right)\sin \Big[\Big(2n+1 \Big)\omega t \Big]
\label{2eqg}
\end{equation}
\begin{equation}
\cos \Big(\beta_0\sin\omega t \Big)=J_0\left(\beta_0\right)+2\sum_{n=1}^{\infty}J_{2n}\left(\beta_0\right)\cos2n\omega t
\label{2eqh}
\end{equation}
 \end{mathletters}
 \smallskip
 
 \noindent The $\sin\beta$ expansion in Equation \ref{2eqg} generates odd harmonics of the PEM reference frequency, while the $\cos\beta$ expansion in Equation \ref{2eqh} generates even harmonics. When propagating  incident light through POLISH as per Equation \ref{2eqq}, one can see that the linear Stokes parameters $Q/I$ and $U/I$ will be modulated at even harmonics of the PEM modulation, while the circular Stokes parameter $V/I$ will be modulated at odd harmonics. Thus, we choose to set our lock-in amplifiers to record the second harmonic of modulated intensity, which is

 \begin{eqnarray}
 \frac{2}{E} I_{\rm obs} = I \! \! \! \! & \pm & \! \! \! \! \! \left[\cos^2 2\theta \cos 2\phi \mp \frac{1}{2} \sin 4\theta \sin 2\phi + \left( \sin^2 2\theta \cos 2\phi + \frac{1}{2} \sin 4\theta \sin 2\phi \right) J_0 \left(\beta_0 \right) \right] Q \nonumber  \label{2eqaa} \\
 & \pm & \! \! \! \! \! \left[ \sin^2 2\theta \cos 2\phi + \frac{1}{2} \sin 4\theta \sin 2\phi\right] \Big[2J_2 \left( \beta_0\right) \cos 2 \omega t \Big] Q \\
 & \pm & \! \! \! \! \! \left[\cos^2 2\theta \sin 2\phi + \frac{1}{2} \sin 4\theta \cos 2\phi + \left( \sin^2 2\theta \sin 2\phi \mp \frac{1}{2} \sin 4\theta \cos 2\phi \right) J_0 \left(\beta_0 \right) \right] U \nonumber \\
 & \pm & \! \! \! \! \! \left[ \sin^2 2\theta \sin 2\phi \mp \frac{1}{2} \sin 4\theta \cos 2\phi\right] \Big[2J_2 \left( \beta_0\right) \cos 2 \omega t \Big] U \nonumber 
\end{eqnarray}
\smallskip

\noindent The lock-in amplifiers record the RMS value of the AC component of the intensity, given by $R$. The amplitude of the AC signal is therefore

{\small
 \begin{equation}
 R\sqrt{2} =  \frac{E}{2} \left\{ \pm 2J_2 \left( \beta_0\right) \left[ \left( \sin^2 2\theta \cos 2\phi + \frac{1}{2} \sin 4\theta \sin 2\phi\right) Q + \left( \sin^2 2\theta \sin 2\phi \mp \frac{1}{2} \sin 4\theta \cos 2\phi\right) U \right] \right\}
 \label{2eqr}
 \end{equation}}
 
 \noindent The mean intensity, or DC level, is given by
 
 {\footnotesize
 \begin{eqnarray}
  \frac{2}{E} {\rm DC} = I \! \! \! \!  & \pm & \! \! \! \! \! \left[\cos^2 2\theta \cos 2\phi \mp \frac{1}{2} \sin 4\theta \sin 2\phi + J_0 \left(\beta_0 \right) \left( \sin^2 2\theta \cos 2\phi + \frac{1}{2} \sin 4\theta \sin 2\phi \right) \right] Q \label{2eqs} \\
  & \pm & \! \! \! \! \! \left[\cos^2 2\theta \sin 2\phi + \frac{1}{2} \sin 4\theta \cos 2\phi + J_0 \left(\beta_0 \right) \left( \sin^2 2\theta \sin 2\phi \mp \frac{1}{2} \sin 4\theta \cos 2\phi \right) \right] U \nonumber
 \end{eqnarray}}
\smallskip
 
 \noindent For $\theta = \pm 45^\circ$, Equations \ref{2eqr} and \ref{2eqs} reduce to
 
\begin{mathletters}
 \begin{equation}
 R\sqrt{2} =  \pm \frac{E}{2} \Big[ 2J_2 \left( \beta_0\right) \Big( Q \cos 2\phi + U \sin 2\phi \Big) \Big]
 \end{equation}
 
 \begin{equation}
 {\rm DC} = \frac{E}{2} \Big[I \pm Q J_0 \left(\beta_0 \right) \cos 2\phi \pm U J_0 \left(\beta_0 \right) \sin 2\phi  \Big]
 \end{equation}
 \end{mathletters}
 \smallskip
 
\noindent Two integrations with the Cassegrain ring rotated $\Delta\phi = 45^\circ$ apart are therefore required to determine both linear Stokes parameters $Q/I$ and $U/I$.\\

For Cassegrain ring angle $\phi = 0^\circ$, the normalized polarization in terms of the observables $R$ and DC is given by

 \begin{equation}
 \frac{Q}{I} = \frac{R \sqrt{2}}{2J_2 \left( \beta_0\right) {\rm DC} - J_0 \left( \beta_0\right) R \sqrt{2}}
 \label{2eqt}
 \end{equation}
\smallskip

\noindent It can be seen that choice of $\beta_0 = 2.4048$ radians, the first zero of $J_0(z)$, ensures that polarization is proportional to the ratio of $R$ and DC. In this case, Equation \ref{2eqt} reduces to Equation \ref{2eqj}, and the PEM efficiency is $E_{\rm PEM} \equiv 2J_2 \left( \beta_0\right) = 86.4\%$. However, choosing halfwave retardance, $\beta_0 = \pi$ radians, increases PEM efficiency to $E_{\rm PEM} = 97.1\%$ at the expense of additional calibration. That is, the DC level is a function of source polarization for halfwave retardance, but it is independent for retardance of $\beta_0 = 2.4048$ radians. Since we plan to observe unpolarized well as strongly polarized sources, we choose a peak retardance of $\beta_0 = 2.4048$ radians in our observations. \\

  \section{Noise}  
 
 Given pre-gain signal current $i_0$ and pre-gain dark current $i_d$, the number of pre-gain photoelectrons during an integration of duration $t_{\rm{AC}}$, and the shot noise on this quantity, will be given by 
 
 \begin{mathletters}
 \begin{equation}
 n = \frac{\left(i_0 + i_d\right) t_{\rm{AC}} }{e}
 \end{equation}
 \begin{equation}
 \sigma_n = \left[\frac{\left(i_0 + i_d\right) t_{\rm{AC}} }{e }\right]^{1/2}
 \end{equation}
 \end{mathletters}
 \smallskip
 
 \noindent The post-gain shot noise current is
 
 \begin{equation}
 \sigma'_{i} = G\left[\frac{2e B F \left(i_0 + i_d\right)}{t_{\rm{AC}}} \right]^{1/2}
 \end{equation}
 \smallskip
 
 \noindent where $B$ is detector bandwidth, $G$ is detector gain, and $F \approx G^x$ is the gain noise factor (post-gain quantities are primed). This factor arises because the gain process itself has statistical fluctuations. The excess noise factor, $x$, is a constant. \\
 
 Since voltage is measured at the output of the detectors, we convert signal and dark current to signal and dark voltage. The pre-gain signal and dark current are $i_0 = {\rm DC} / G T_A$ and $i_d = i'_d / G$, where $T_A$ is the amplifier transimpedance in V/A and the post-gain dark current is $i'_d$. Output noise voltage is related to noise current by $\sigma'_{v} = T_A \sigma'_{i}$, so
 
 \begin{equation}
 \sigma'_{v} = \left[\frac{ 2 e B G^{1+x} T_A ({\rm DC} + i'_d T_A)}{t_{\rm{AC}}} \right] ^{1/2}
 \label{2eqao}
 \end{equation}
 \smallskip
 
 \noindent This is the expected voltage noise at the output of imperfect detectors, where $x \neq 0$. The bandwidth $B$ must be chosen with care. Photon shot noise is white noise, but it cannot have infinite bandwidth because integrated power would also be infinite. The maximum frequency at which photon shot noise can occur is the count rate of noise photons. Thus, we determine maximum bandwidth by the square root of the number of detected photons,
  
  \begin{equation}
  B_{\rm max} = \left(\frac{{\rm DC}}{e G T_A}\right)^{1/2}
  \label{2eqav}
  \end{equation}
 \smallskip
  
  \noindent The detectors have bandwidths between about 100 kHz and 200 kHz. However, the lock-in amplifiers only admit noise in a bandwidth of $B_{\rm AC} \approx 2.6$ Hz, while the voltmeters have bandwidth of $B_{\rm DC} \approx 1$ MHz. Thus, the bandwidth of shot noise on $X$ and $Y$ will be ${\rm min} (B_{\rm max}, B_{\rm detector}, B_{\rm AC}) = B_{\rm AC}$ for all stars. The bandwidth of shot noise on DC will be ${\rm min} (B_{\rm max}, B_{\rm detector}, B_{\rm DC})$, which will depend on stellar intensity. \\
  
  Using Equation \ref{2eqao}, we can determine photon shot noise ($x = 0$ and $i'_d = 0$) and detector noise ($x \neq 0$ and $i'_d \neq 0$) on $X$, $Y$, and DC. We can also compare these values to the observed fluctuations during each measurement.
  
  \begin{mathletters}
  \begin{equation}
  \left[ \begin{array}{c} \sigma_X \\ \sigma_Y \\ \sigma_{\rm{DC}} \end{array} \right]_{\rm{shot}} = \gamma_0 \left[ \begin{array}{c} B_{\rm AC}  \\  B_{\rm AC} \\  {\rm  min} \left(B_{\rm max},B_{\rm detector}\right) \end{array} \right]^{1/2}
  \label{2eqaq}
  \end{equation}
  \begin{equation}
  \left[ \begin{array}{c} \sigma_X \\ \sigma_Y \\ \sigma_{\rm{DC}} \end{array} \right]_{\rm{detector}} = \gamma  \left[ \begin{array}{c} B_{\rm AC}  \\  B_{\rm AC} \\  {\rm  min} \left(B_{\rm max},B_{\rm detector}\right) \end{array} \right]^{1/2}
  \label{2eqar}
  \end{equation}
  \begin{equation}
  \left[ \begin{array}{c} \sigma_X \\ \sigma_Y \\ \sigma_{\rm{DC}} \end{array} \right]_{\rm{obs}} =  \left[ \begin{array}{c} {\rm std}(X_{\rm{src}})^2 + {\rm std}(X_{\rm{sky}})^2  \\  {\rm std}(Y_{\rm{src}})^2 + {\rm std}(Y_{\rm{sky}})^2\\  {\rm std}({\rm DC}_{\rm{src}})^2 + {\rm std}({\rm DC}_{\rm{sky}})^2 \end{array} \right]^{1/2}
  \label{2eqas}
  \end{equation}
 \end{mathletters}
 \smallskip
 
\noindent Here, $\gamma_0 \equiv 2 e G T_A {\rm DC}$, $\gamma \equiv 2 e G^{1 + x} T_A ({\rm DC} + i'_d T_A)$, and ``std" indicates the sample standard deviation. We now relate this quantity to fluctuations in the observables $X$, $Y$, and DC. By propagating error through Equation \ref{2eqj}, we find polarimetric uncertainty of a measurement to be related to uncertainty in $X$, $Y$, and DC according to
 
 \begin{equation}
 \sigma_P = \frac{\sqrt{2}}{E_{\rm{PEM}} {\rm DC}} \left\{ \frac{1}{t_{\rm{AC}}} \left[ \frac{X^2 \sigma_X^2 + Y^2 \sigma_Y^2}{X^2 + Y^2} +\frac{1}{2} \Big( E_{\rm{PEM}} P \sigma_{\rm{DC}} \Big)^2 \right] \right\} ^{1/2}
 \label{2eqap}
 \end{equation}
 \smallskip
 
 \noindent Finally, inserting Equations \ref{2eqaq} through \ref{2eqas} into Equation \ref{2eqap} yields the uncertainty in polarization from photon shot noise, detector noise, and observed fluctuations, given as Equations \ref{2eqau} through \ref{2eqat}. \\

 \end{document}